\documentclass[a4paper,11pt]{article}
\PassOptionsToPackage{table}{xcolor}

\usepackage{jheppub} \usepackage[T1]{fontenc} 

\usepackage[all]{xy}

\usepackage[percent]{overpic}
\usepackage{cancel}
\usepackage{slashed}
\usepackage{wrapfig}
\usepackage{tabu}
\usepackage{diagbox}
\usepackage{mathrsfs,amsmath,amssymb,amsthm,amsfonts,tikz,graphicx,accents,hyperref, color}
\usepackage{dsfont,epiolmec, latexsym, stmaryrd, comment}
\usepackage{slashed,ccaption}
\usepackage{mathrsfs, calligra}
\usepackage{leftidx}
\usepackage{import}
\usepackage{multirow}
\usepackage{amsfonts}
\usepackage{pifont}
\usepackage{tabularx}
\usepackage[utf8]{inputenc}
\usetikzlibrary{intersections,calc}
\usepackage{tikz-3dplot}
\usepackage{ifthen}
\usetikzlibrary{arrows}
\usepackage{amsmath}
\usepackage{xcolor}

\usepackage{array}

\hypersetup{ linktoc=all,
    colorlinks, linkcolor={palatinateblue},
    citecolor={brightpink}, urlcolor={amaranth}
}

\graphicspath{{Images/}}

\renewcommand{\d}[1]{\ensuremath{\operatorname{d}\!{#1}}}

\def\sideremark#1{\ifvmode\leavevmode\fi\vadjust{\vbox to0pt{\vss% the remark
 \hbox to 0pt{\hskip\hsize\hskip1em%                          will appear only
 \vbox{\hsize2cm\tiny\raggedright\pretolerance10000%          on the side
 \noindent #1\hfill}\hss}\vbox to8pt{\vfil}\vss}}}%
\DeclareSymbolFont{extraup}{U}{zavm}{m}{n}
\DeclareMathSymbol{\varheart}{\mathalpha}{extraup}{86}
\DeclareMathSymbol{\vardiamond}{\mathalpha}{extraup}{87}
\makeatletter
\renewcommand*{\@fnsymbol}[1]{\ensuremath{\ifcase#1\or \clubsuit \or \vardiamond \or \varheart\or
    \spadesuit\or \mathparagraph\or \|\or **\or \dagger\dagger
    \or \ddagger\ddagger \else\@ctrerr\fi}}
\makeatother

\definecolor{rosy}{RGB}{230,235,252}
\definecolor{myframetitle}{RGB}{90,89,170}
\definecolor{myblocktitle}{RGB}{140,185,249}
\definecolor{mytitle}{RGB}{10,80,26}

\definecolor{darkgreen}{RGB}{27,130,45}
\definecolor{darkblue}{rgb}{0,0,0.3}
\definecolor{darkred}{rgb}{0.7,0,0}

\definecolor{light gray}{RGB}{220,220,220}
\definecolor{dark purple}{RGB}{108,0,217}
\definecolor{pink}{RGB}{190,20,100}
\definecolor{orang}{RGB}{193,63,0}
\definecolor{green}{RGB}{11,98,17}
\definecolor{darkpink}{RGB}{153,0,76}
\definecolor{bluegreen}{RGB}{0,102,102}
\definecolor{greenlagan}{RGB}{0,102,0}
\definecolor{redgreen}{RGB}{102,102,0}
\definecolor{Redgreen}{RGB}{153,76,0}
\definecolor{vividviolet}{rgb}{0.62, 0.0, 1.0}
\definecolor{amaranth}{rgb}{0.9, 0.17, 0.31}
\definecolor{palatinateblue}{rgb}{0.15, 0.23, 0.89}
\definecolor{brightpink}{rgb}{1.0, 0.0, 0.5}
\definecolor{cornflowerblue}{rgb}{0.39, 0.58, 0.93}
\definecolor{deepcarminepink}{rgb}{0.94, 0.19, 0.22}
\definecolor{radicalred}{rgb}{1.0, 0.21, 0.37}
\newcommand{\be}{\begin{equation}}
\newcommand{\ee}{\end{equation}}
\newcommand{\bea}{\begin{eqnarray}}
\newcommand{\eea}{\end{eqnarray}}
\DeclareFontFamily{OT1}{rsfs}{}

\DeclareFontShape{OT1}{rsfs}{m}{n}{ <-7> rsfs5 <7-10> rsfs7 <10->rsfs10}{} 

\DeclareMathAlphabet{\mycal}{OT1}{rsfs}{m}{n}
\makeatletter \@addtoreset{equation}{section}

\begin{document}
\preprint{IPM/P-2020/043}
\rightline{BIMSA/P-1}

\newcommand{\mytitle}{\LARGE{Symmetries at Null Boundaries:}\\ 
{\Large{Two and Three Dimensional Gravity Cases}}}
\title{\center{\textbf{\mytitle}}}

\author[a]{H.~Adami}
%\author[b]{, D.~Grumiller}
\author[a]{, M.M.~Sheikh-Jabbari}
\author[b]{, V.~Taghiloo}
\author[c]{, H.~Yavartanoo}
\author[d]{and C.~Zwikel}
\affiliation{$^a$ School of Physics, Institute for Research in Fundamental
Sciences (IPM),\\ P.O.Box 19395-5531, Tehran, Iran}
%\affiliation{$^c$ The Abdus Salam ICTP, Strada Costiera 11, Trieste, Italy}
\affiliation{$^b$ Department of Physics, Institute for Advanced Studies in Basic Sciences (IASBS),
P.O. Box 45137-66731, Zanjan, Iran}
\affiliation{$^c$ Beijing Institute of Mathematical Sciences and Applications (BIMSA), Huairou District, Beijing 101408, P. R. China}
\affiliation{$^b$ Institute for Theoretical Physics, TU Wien, Wiedner Hauptstrasse 8--10/136, A-1040 Vienna, Austria}

\emailAdd{hamed.adami@ipm.ir,  jabbari@theory.ipm.ac.ir, \\ v.taghiloo@iasbs.ac.ir, yavar@itp.ac.cn, celine.zwikel@tuwien.ac.at}

\abstract{We carry out in full generality and without fixing specific boundary conditions, the symmetry and charge analysis near a generic null surface for two and three dimensional ($2d$ and $3d$) gravity theories. In $2d$ and $3d$ there are respectively two and three charges which are generic functions over the codimension one null surface. The integrability of charges and their algebra depend on the state-dependence of symmetry generators which is a priori not specified.  We establish the existence of infinitely many choices  that render the surface charges integrable.  We show that there is a choice, the ``fundamental basis'', where the null boundary symmetry algebra is the Heisenberg$\oplus$Diff($d-2$) algebra. We expect this  result to be true for $d>3$ when there is no Bondi news through the null surface.}
\maketitle

%%%%%%%%%%%%%%%%%%%%%%%%%%%%%%%%%%%%%%%%%%%%%%%%%%%%%%%%%%%%%5
\section{Introduction}\label{sec:1}
%%%%%%%%%%%%%%%%%%%%%%%%%%%%%%%%%%%%%%%%%%%%%%%%%%%%%%%%%%%%%5
A standard way to describe a $d$-dimensional field theory at {the} classical level is through an action which is the integral of a Lagrangian over a $d$-dimensional manifold ${\cal M}$. For physical theories, ${\cal M}$ is typically a Lorentzian manifold which may have a codimension one boundary $\partial {\cal M}$. {The} variational principle can then be used to derive equations of motion which are typically second order differential equations over ${\cal M}$. These equations are well-posed if their solutions can be fully determined by specifying the boundary data which e.g. 
can be the Cauchy data over a constant time slice in ${\cal M}$.

Gravity theories in generic $d$ dimensions have two important features: (1) Background independence, meaning that the metric on ${\cal M}$ is a solution to the same theory and not a priori fixed or given. In particular, to completely specify it we need to provide the boundary data.  (2) Diffeomorphism invariance, meaning that the metric can only be determined up to generic coordinate transformations. This implies  that among $d(d+1)/2$ components of the $d$-dimensional metric, $d$ number of them can be removed by an appropriate choice of coordinates. Out of the remaining $d(d-1)/2$, only $d(d-3)/2$ correspond to propagating gravitons\footnote{Here we are implicitly assuming an Einstein gravity theory which has only a massless spin 2 state as its propagating degree of freedom.} and $d$ are free to choose. This latter will be fixed through the boundary data which are functions over codimension one surfaces, i.e. we need $d$ functions of $d-1$ variables to specify the boundary data \cite{Grumiller:2020vvv}. 

Discussions of the previous two paragraphs  raise  a crucial point to be addressed and understood:  $\partial{\cal M}$ is a $d-1$ dimensional surface and a generic $d$-dimensional diffeomorphism may non-trivially act on the boundary data and hence move us within the space of solutions determined by the boundary data. This means that a part of the general local coordinate transformations which act non-trivially on the boundary data can become ``physical'' as they change a solution to another physically distinguishable solution. We hence need to refine the equivalence principle, which leads to diffeomorphism invariance, in {the} presence of boundaries \cite{Sheikh-Jabbari:2016lzm}. This argument already tells us that this (potentially)  ``physical diffeos'' should be a measure zero subset of $d$-dimensional diffeos which act on the {codimension} one boundary surface. That is, they can at most be $d$ functions of $d-1$ variables parametrising the boundary \cite{Grumiller:2020vvv}. These physical diffeos are indeed labelling the boundary degrees of freedom (b.d.o.f.).  In {the} presence of boundaries, especially if they are timelike or null, the system besides the $d(d-3)/2$ bulk gravitons has a maximum number of $d$ b.d.o.f. They can interact among themselves and also with the bulk d.o.f. The details of these interactions are of course determined once we fix the boundary conditions.

{Our general goal is to develop a systematic treatment of the b.d.o.f. and their interactions with the bulk d.o.f.}
The key questions toward this goal are 
\begin{enumerate}
    \item Whether and how this maximal physical diffeos and/or b.d.o.f. can be realised; 
    \item What are the possible choices for boundary conditions, what is the guiding principle; 
    \item How does this fix the interactions mentioned above.
\end{enumerate}
These questions have of course been under intense study since the seminal work of BMS \cite{Bondi:1962, Sachs:1962} and have gained a boost in the last decade, e.g. see \cite{strominger:2017zoo} and references therein.  
{In this work we focus on the first question. The two other questions will be briefly discussed in the conclusion section and get a full treatment in upcoming publications.} 

We can systematically analyse this question by fixing/choosing a codimension one boundary surface  using $d$ diffeomorphisms. We then remain  with $d$ ``residual'' diffeos which only act on the chosen $d-1$ dimensional ``boundary''. To specify which of these diffeos are trivial or not, we can employ one of the standard methods of associating (conserved) charges to the diffeos, e.g. the covariant phase space method \cite{Lee:1990nz,Iyer:1994ys,Compere:2018aar}; if the charge is finite and non-zero the diffeo is nontrivial {or physical}.  This method, however,  provides us with {the} variation of the charges and not the charges themselves. There is then another non-trivial step to check if the charge is integrable over the space of solutions discussed above. If integrable,  one can then define the charge. The same covariant phase space formulation also implies the algebra of these integrable charges is the same as the Lie algebra of the symmetry generating diffeos, up to possible central charges, e.g. see \cite{Compere:2018aar}. These charges, if well-defined, can be used to label the b.d.o.f., i.e. b.d.o.f. fall into representations (more precisely, coadjoint orbits) of the charge algebra, see  e.g. \cite{Oblak:2016eij}. 

Classic examples are four or three-dimensional asymptotically flat spacetimes where the boundary is the future null infinity $\boldsymbol{\mathscr{I}^+}$. In these cases, the residual diffeos are respectively,  three and two functions over codimension two surfaces. The charges turn out to be integrable in the absence of the Bondi news \cite{Bondi:1962, Wald:1999wa, Barnich:2011mi} and the algebra  is respectively   BMS$_4$ \cite{Bondi:1962, Sachs:1962} or BMS$_3$ \cite{Barnich:2010eb}. In both cases, the maximal numbers of b.d.o.f is not reached and moreover, the symmetry generators are only functions of codimension two surfaces (rather than codimension one). 
%As classic examples,  for four or three-dimensional asymptotically flat spacetimes one may take the boundary to be the asymptotic null surface $\boldsymbol{\mathscr{I}^+}$. In the literature only a part of the maximal residual diffeos (respectively,  four or three functions over codimension two surfaces, rather than functions over codimension one surface $\boldsymbol{\mathscr{I}^+}$) has been considered for which the charges turn out to be integrable in the absence of the Bondi news \cite{Bondi:1962} and the algebra in this sector is respectively   BMS$_4$ \cite{Bondi:1962, Sachs:1962} or BMS$_3$ \cite{Barnich:2010eb}. 
These considerations can also be applied to timelike (causal) boundaries as in the seminal case of asymptotically AdS$_3$ \cite{Brown:1986nw}. The Brown-Henneaux analysis yields integrable charges for only a subsector of the maximal boundary data which has two functions of a single variable (rather than three functions of two variables). The algebra is two copies of Virasoro algebra at the Brown-Henneaux central charge \cite{Brown:1986nw}.

While it is common to explore  these charge analyses over the asymptotic boundaries of spacetimes, one can take these boundaries to be any codimension one surface. In particular, for the case of black holes, a natural and physically relevant choice is to take the horizon, which is a null surface, to be the boundary over which the boundary data and b.d.o.f. is stored. This choice is physically relevant because the horizon is indeed the boundary of timelike curves (paths of physical observers) outside the horizon. Moreover, one can model what is inside a black hole horizon by replacing the inside region with a membrane, placed at the stretched horizon, as is done in the membrane paradigm \cite{Thorne:1986iy, Parikh:1998mg}. {Furthermore}, near horizon degrees of freedom and symmetries, are the essentials of {the} formulation of the soft hair proposal \cite{Hawking:2016msc} %, Hawking:2016sgy} 
and have been explored in several different publications \cite{Donnay:2015abr, Donnay:2016ejv, Afshar:2016kjj, Afshar:2016wfy, Afshar:2016uax, Mao:2016pwq, Grumiller:2016kcp, Grumiller:2018scv, Ammon:2017vwt, Chandrasekaran:2018aop, Chandrasekaran:2019ewn, Grumiller:2019fmp, Adami:2020amw}. Nonetheless, 
%in these examples  different boundary conditions have been imposed and 
neither realize the maximal set of b.d.o.f. discussed above. 

With the above motivations, in this work, we focus on boundary charges over generic null surfaces in two-dimensional dilaton gravity and three-dimensional Einstein gravity theories. These theories do not have propagating d.o.f. and {thus} they provide a more controlled setup to ask, formulate, and address  questions about boundary charges in full generality. As we will see this setup is still very rich and provides us with results and insights which could be directly generalized to  higher-dimensional cases. In our surface charge analysis, {we realize} the largest set of boundary charges by keeping the boundary conditions free and unfixed. {That is, unlike almost all previous analysis, we do not fix any boundary conditions; fixing boundary conditions generically amounts to a reduction on the phase space governing the b.d.o.f.}

In the $2d$ case these are two sets of charges which are functions of the one variable parametrising the null surface. In the $3d$ case these are three sets of charges which are functions of two variables spanning the null surface, which is a null cylinder. {The symmetry generators (non-trivial diffeos) can be chosen to be field/state-dependent, meaning that they can depend on the functions defining the phase space of solutions.}  One of our main results is that in the $2d$ and $3d$ cases there always exists a basis where the charges are integrable. The algebra of these charges, however, depends on the basis used for symmetry generators: a state-dependent symmetry generator will change the algebra of charges. We discuss some different charge bases and the corresponding algebras.

The organisation  of this paper is as follows. In section \ref{sec:2dgrav} and \ref{sec:3dgrav}, we consider dilation gravity in $2d$ and pure gravity in $3d$ respectively. Based on the most general expansion around a null hypersurface, we compute the null boundary symmetries (NBS) and their charges. We show, by construction, that there exists a family of reparametrisation of the symmetries yielding integrable charges. {For $2d$ case we obtain a Heisenberg algebra and in $3d$,  Heisenberg semidirect sum with Diff$(S^1)$ as the algebra of charges.}   In section \ref{sec:changebasis}, we discuss generic reparametrisations/change of bases and discuss various algebras one can reach for both the $2$ and $3d$ cases. The last section is devoted to concluding remarks and outlook.  In particular, we discuss that the ``fundamental null boundary symmetry''   algebra, Heisenberg $\oplus$ Diff$(d-2)$, that can be always reached in generic dimensions (including $d>3$) in the absence of Bondi news through the null surface. 
In some appendices, we have gathered some useful formulas regarding $2d$, $3d$ gravity theories, and their general solutions and discuss modified bracket method \cite{Barnich:2011mi} to deal with non-integrable charges

%%%%%%%%%%%%%%%%%%%%%%%%%%%%%%%%%%%%%%%%%%%%%%%%%%%%%%%%%%%%%%%%%
\section{Null Boundary Symmetry (NBS) Algebra, \emph{2d} Dilaton Gravity Case}\label{sec:2dgrav}
%%%%%%%%%%%%%%%%%%%%%%%%%%%%%%%%%%%%%%%%%%%%%%%%%%%%%%%%%%%%%%%%%%%%%%%%%%%%%%%%%%%%%%%%%%%%%%%%%%%%%%%%%%%%%%%%%%%%%%%%%%%%%%%%%%%%

Two-dimensional spacetime is the lowest dimension for which one can consider gravity. However, the Einstein-Hilbert action is purely topological in $2d$. One simple way to get a bulk action and therefore equations of motion is to add a scalar field, the dilaton \cite{Jackiw:1984, Teitelboim:1984, Callan:1992rs}. This set-up has no propagating degrees of freedom nevertheless very interesting features, e.g. see \cite{Brown:1988am, Grumiller:2002nm} and references therein. In this work, we focus on the computations of the charges on a null hypersurface, see  \cite{Grumiller:2015vaa, Grumiller:2017qao} for earlier analysis of asymptotic symmetry analysis on AdS$_2$. 

The dilaton-gravity action is \cite{Jackiw:1984, Teitelboim:1984}
\be 
S_G=\frac{1}{16\pi G}\int \textrm{d} ^2 x \sqrt{-g}\left(\phi^2 R-\lambda(\partial\phi)^2-V(\phi)\right)\,.
\ee 
However, the kinetic term of dilaton can be absorbed by a field redefinition
\be
g\to \phi^{-\frac{\lambda}{2}}g,\quad\Rightarrow \quad \lambda\to 0, \quad V\to \phi^{-\frac{\lambda}{2}} V\,.
\ee
So we consider the following action \cite{Grumiller:2002nm} \footnote{We don't consider any extra matter field in the system nor discuss the boundary terms.}
\be \label{action2d}
S=\frac{1}{16\pi G}\int \textrm{d} ^2 x \sqrt{-g}\left(\Phi R-U(\Phi)\right)\,. %+S_{matter}+S_{b'dry}
\ee 
The above theory with different potentials $U(\Phi)$ have been considered in the literature, e.g. $U(\Phi)=\Phi^k$ type potential may arise from various reduction of higher-dimensional theories to two dimensions. See \cite{Grumiller:2002nm, Grumiller:2017qao} for more discussions and references. For our analysis below, where we consider an expansion around a null surface, however, as long as the potential is not a constant its explicit form  does not matter.

The most general solution to the above $2d$ gravity theory has been discussed in the appendix \ref{appen:2d-solutions}. There are two classes of families, constant $\Phi$ and non-constant $\Phi$ solutions. The former, however, has always vanishing charges. So, hereafter we only consider {a} non-constant dilaton family of solutions. 

For the charge analysis,  we only need to have the leading behaviour of the fields near the $r=0$ surface, which we take it to be null,
\be\label{2d-NH-metric}
\begin{split}
\d s^2 &= 2\eta(v) \d r \d v- 2r \eta(v)  F_0(v)\d v^2 +{\cal O}(r^2)\,\\
\Phi &= \Phi_0 (v)+ {\cal O}(r)
\end{split}
\ee
The field equations to first order in $r$ relate the three functions $\eta, F_0,\Phi_0$, 
\begin{equation}
       % &  2 F_1+ U_0 =0 \, , \\
        - \Phi_0'' + \Phi_0' \left( F_0 +  \frac{\eta'}{\eta} \right)=0 \, , 
       % & 2 \Phi_0' + \Phi_1 F_0 + \eta U_0 =0 \, ,
\end{equation}
{For our charge analysis, we only focus on the non-constant $\Phi_0$, $\Phi_0'\neq0$, case.\footnote{{In the special case that $\Phi_0'=0$, $F_0, \eta$ are the two arbitrary functions describing the solution near the null boundary $r=0$.}} The system is then described  by $\eta, \Phi_0$ which have arbitrary $v$ dependence, as $F_0$ is given by}
\begin{equation}\label{F0}
F_0= \frac{\Phi''_0}{\Phi'_0}-\frac{\eta'}{\eta} \,.
\end{equation}
In the special case that $\Phi_0'=0$, $F_0, \eta$ are the two arbitrary functions describing the solution near the null boundary $r=0$.
As discussed in appendix \ref{appen:2d-solutions}, the solutions admit
\begin{equation}
    \zeta= \epsilon^{\mu \nu} \partial_\mu \Phi \partial_\nu
\end{equation}
as a Killing vector, where $\epsilon^{\mu \nu}$ is Levi-Civita tensor in 2$d$ \cite{Gegenberg:1994pv}. This vector is normal to the charge computing surface which is along  $\partial_\mu\Phi$. Since $|\zeta|^2=|\d\Phi|^2$, the Killing vector field $\zeta$ is null on $r=0$ when $\partial_v\Phi_0\partial_r\Phi|_{r=0}=0$.
%\hnote{For non-constant $\Phi_0$ which is an essential assumption in case II, one can simply show that the leading term of Eq.~\eqref{FF-case-II} is of the form \eqref{F0}.}

\subsection{NBS generating vector fields} 
The vector fields that preserve the form of solution \eqref{2d-NH-metric} and keep $r=0$ a null surface are
\begin{equation}\label{2d-NHKV-flat}
\xi=T\partial_{v}-r(W-T')\partial_{r}+ \mathcal{O}(r^2) \, ,
\end{equation}
where $T,W$ are arbitrary  functions of $v$. The Lie bracket of two vectors of the form \eqref{2d-NHKV-flat} is
\begin{equation}\label{2d-algebra}
\begin{split}
[\xi(T_{1},W_{1}),\xi(T_{2},W_{2})]&=\xi(T_{12},W_{12})\cr
    T_{12}=(T_{1}T_{2}'-T_{2}T_{1}')\,&,\qquad 
    W_{12}=  (T_{1}W_{2}'-T_{2}W_{1}')\,.
\end{split}
\ee
If we assume that $T,W$ are meromorphic functions of $v$, i.e. if they admit a Laurent expansion,\footnote{{Assuming smooth diffeos, one should consider Taylor expansion, i.e. the sums in \eqref{T-W-Laurent} should be limited to non-negative integers $n=0,1,2,\cdots$. Allowing for negative $n$ amounts to considering diffeos which have poles in $v$. This is somehow like the superrotation charge of  BMS$_4$ [12, 38]}}
\begin{equation}\label{T-W-Laurent}
    T=-\sum_{n\in Z} \tau_n v^{n+1},\qquad W= \sum_{n\in Z} \omega_n v^{n+1},
\end{equation}
\begin{equation}
    T_n=-v^{n+1} \partial_v - (n+1)\, r\, v^n \partial_r,\qquad W_n=- r\, v^{n+1}\partial_r,
\end{equation}
then $T_n, W_n$ satisfy a BMS$_3$ algebra \cite{Barnich:2006av,Barnich:2010eb},
\begin{equation}\label{BMS3-2d-KVA}
    [T_m, T_n]= (m-n)\,T_{m+n}\, , \quad [T_m, W_n]= (m-n)\,W_{m+n},\quad [W_m, W_n]=0.
\end{equation}
It is notable that \eqref{2d-NHKV-flat} is one-dimensional Diff $\oplus$ Weyl algebra, in which the Weyl scaling corresponds to the BMS$_3$ supertranslations. 
%We will return to this BMS$_3$ \tcb{in the appendix \ref{BT-MB-Appendix}.} 

Under a transformation generated by \eqref{2d-NHKV-flat}, the metric and the dilaton become
\be g_{\mu\nu}[\eta, \Phi_0] \to g_{\mu\nu}[\eta+\delta_\xi \eta, \Phi_0+\delta_\xi \Phi_0] 
 \ee 
where
%\snote{Work out field variations}
\begin{equation}\label{xi-trans}
    \delta_\xi \eta= \eta' \, T + 2 \,\eta \, T'-\eta\ W,\qquad 
\delta_\xi \Phi_0=  \Phi_0'\ T\,.
\end{equation}
%%%%%%%%%%%%%%%%%%%%%%%%%%%%%%%%%%%%%%%%%%%%%%%%%%%%%%%%
\subsection{Surface charges}\label{sec:charge-2d-case}
%%%%%%%%%%%%%%%%%%%%%%%%%%%%%%%%%%%%%%%%%%%%%%%%%%%%%%%%%

Using the Iyer-Wald presymplectic form \cite{Iyer:1994ys}, one can compute the charge variation associated with the above transformations.  Straightforward analysis yields
\begin{equation}
   \slashed{\delta} Q_\xi = %\tcr{{\Phi} \mathcal{Q}_\xi^{v r}}
   - \frac{\sqrt{-g}}{8 \pi G} \left( \delta{\Phi} \, \nabla^{[v} \xi^{r]} - \,\xi^{[v} h^{r]}_{\lambda} \nabla^{\lambda} {\Phi} +2 \xi^{[v} \nabla^{r]} \delta{\Phi}  \right),
\end{equation}
where $h_{\mu\nu}=\delta g_{\mu\nu}$ denotes metric variations and $\slashed{\delta}$ indicates that the charge is not necessarily integrable. We note that 
the charge variation is coming from the part proportional to $\delta\Phi$ and the part  not involving variation of $\Phi$ does not contribute to the charge in the $2d$ case for our background metrics \eqref{2d-NH-metric}. The  charges associated with the near null hypersurface Killing vectors \eqref{2d-NHKV-flat} for the metric \eqref{2d-NH-metric}  computed at $r=0$ are  
\bea\label{Charge-variation-01}
16\pi G \slashed{\delta} Q_{\xi}&=& W\delta\Phi_0+T\left(-\Gamma \delta\Phi_0 -2\delta \Phi_0'+\frac{\Phi_0'\delta\eta}{\eta}\right) 
\eea
where
\begin{equation}\label{P-G-01}
\Gamma:= - \frac{2 \, \Phi_0''}{\Phi_0'}+\frac{\eta'}{\eta} \,, \qquad
    \delta_\xi \Gamma = \left(\Gamma \, T \right)' - W' \, .
\end{equation}
If the symmetry generators are state independent $\delta T =\delta W=0$, the {charges} are not integrable on the phase space. Moreover, they are functions of $v$ and their evolution is not constrained by the equations of motion. Non-inegrability of the charge for non-constant dilaton cases was also reported in \cite{Grumiller:2015vaa, Grumiller:2017qao} where asymptotic symmetries on AdS$_2$ was analysed. One method to deal with the non-intergable charges is to use the Barnich-Troessaert modified bracket \cite{Barnich:2011ct, Barnich:2011mi} (see also \cite{Adami:2020amw} for more discussions) to extract out the integrable part of the non-integrable charges. That is, to write  $\slashed{\delta} Q_{\xi}=\delta Q^I_\xi+{\cal F}$ where $Q^I$ is the integrbale charge and $F$ is a non-zero ``flux''. As we show in the appendix \ref{BT-MB-Appendix} the algebra of $Q^I$ is a centerless BMS$_3$ algebra, the same algebra as the symmetry generators \eqref{BMS3-2d-KVA}. In the next part, however, we explore a different line and find a basis in which the charges are integrable.

%%%%%%%%%%%%%%%%%%%%%%%%%%%%%%%%%%%%%%%%%%%%%%%%%%%%%%%%%%%%%%%%%%%
\subsection{Integrable basis for the charges}\label{sec:integrabe-2d-basis}
%%%%%%%%%%%%%%%%%%%%%%%%%%%%%%%%%%%%%%%%%%%%%%%%%%%%%%%%%%%%%%%%%%%%
%
As discussed in \cite{Grumiller:2019fmp}, we may choose which combination of the symmetry generators are assigned to have vanishing variations. That is, one may consider ``change of basis'' by taking a linear combination of the symmetry generators, $W,T$ with possibly field-dependent coefficients to define a new basis. In particular, let us consider
\begin{equation}\label{2dInt-basis-generators}
    \hat W=W-\Gamma\, T\,,\quad \hat T=\left(\frac{\eta}{(\Phi_0')^2}\right)^{s}{\Phi_0'}\ {T}\, 
\end{equation}
where $\Gamma$ is defined in \eqref{P-G-01} and $s$ is an arbitrary real number. In the new basis we take $\hat{W}$ and $\hat{T}$ to be two field-independent functions and to have zero variations over the phase space.

{The field-dependence of the symmetry generators has to be taken account when one computes the symmetry algebra}
This is done systematically through the adjusted Lie bracket is defined as \cite{Barnich:2010eb, Compere:2015knw},
\be [\xi_1,\xi_2]_{_{\text{adj. bracket}}}=[\xi_1, \xi_2]-\hat\delta_{\xi_1}\xi_2+\hat\delta_{\xi_2}\xi_1\ee where $\hat\delta$ is coming from variation of fields $T, W$ in $\xi_1,\xi_2$. 
Their algebra is 
\begin{equation}\label{NHKV-algebra-2d}
    [\xi(\hat W_1, \hat T_1), \xi(\hat W_2, \hat T_2)]_{_{\text{adj. bracket}}}=\xi\left(0,s(\hat T_1\hat W_2-\hat T_2\hat W_1)\right)\,.
\end{equation}

The charge variation \eqref{Charge-variation-01} in this new basis takes the form
\begin{equation}\label{charge-vartiation-02}
     \delta Q_\xi=\frac1{16\pi G}\left( \hat W \, \delta \Phi_0+\hat T\, \delta \mathcal{P}^{(s)}\right)
\end{equation}
where
\begin{equation}\label{E-01}
    \mathcal{P}^{(s)}=\left\{\begin{array}{cc}-\frac1s\left( \frac{(\Phi'_0)^2}{\eta}\right)^{s} \,, &\text{ for } s\neq0\\ \ \ & \ \\
     - \ln \left( \frac{(\Phi'_0)^2}{\eta}\right)\,, &\text{ for } s=0\end{array}\right. \,.
\end{equation}
The charges $\mathcal{P}^{(s)}, \Phi_0$ are clearly integrable once we take $ \hat W, \hat T$ to be field-independent in the new basis. This is not a surprise, as in the absence of propagating degrees of freedom in $2d$, one can expects to be able to find a new basis in which flux vanishes. In section \ref{sec:changebasis} we provide a general discussion on the existence of integrable basis.  

The transformation law of the fields is given by 
\begin{equation}
     \begin{cases} \delta_\xi\mathcal{P}^{(s)}= s\,\hat{W}\,\mathcal{P}^{(s)} ,\qquad \delta_\xi \Phi_0=-s\mathcal{P}^{(s)}\hat{ T}\, & \qquad s\neq0\\ \delta_\xi\mathcal{P}^{(0)}= -\hat{W}, \hspace{2 cm} \delta_{\xi}\Phi_{0}=\hat{T} & \qquad s=0\,,\end{cases}
\end{equation}

yielding the charge algebra, 
\begin{equation}\label{2d-Phi-Xi-algebra}
    \begin{split}
       i \{ \Phi_0 (v), \mathcal{P}^{(s)} (v) \} = &{16\pi G}i\left(-s\,\mathcal{P}^{(s)}(v)+ \,\delta_{s,0} \right)\,  \\
       i\{ \Phi_0 (v), \Phi_0(v) \} = &\ i\{ \mathcal{P}^{(r)}  (v), \mathcal{P}^{(s)} (v) \}=0 \, .
    \end{split}
\end{equation}

Let's start by considering the case $s=0$. Performing ``quantisation'' of the algebra by replacing the Poisson brackets with commutators, $i\{ , \}\to [ , ]$, one has 
\begin{equation}\label{C-R-01}
\begin{split}
&[\Phi_{0}(v),\mathcal{P}^{(0)}(v)]=16\pi G\ {i} \\ %\delta(v-u)\\
&[\Phi_{0}(v),\Phi_{0}(v)]=[\mathcal{P}^{(0)}(v),\mathcal{P}^{(0)}(v)]=0\,,
\end{split}
\end{equation}
which is the Heisenberg algebra. Therefore in our setup, we have a change of basis where we can reach the Heisenberg algebra  \eqref{C-R-01} with $v$ dependent charges as the  NBS algebra in {the} $2d$ case. In particular, we would like to emphasise two points:
\begin{itemize}
    \item [1] The null boundary is a one-dimensional line parameterised by $v$ and the boundary phase space is labelled by the b.d.o.f $\Phi_{0}(v),\mathcal{P}^{(0)}(v)$. That is, the b.d.o.f. in this case is that of a one-dimensional particle. 
    \item [2] The $v$ dependence of these charges is not fixed by our charge analysis because we did not fix any boundary condition on the null surface $r=0$. Therefore, the boundary Hamiltonian (generator of translations in $v$) which governs the dynamics over the boundary phase space is not determined through our analysis. We shall discuss this point further in section \ref{sec:conclusion}.
\end{itemize}

To close the section, we note that the generic $s\neq 0$ case of \eqref{2d-Phi-Xi-algebra} can be obtained from the Heisenberg algebra through 
\begin{equation}\label{Es-E-1}
\mathcal{P}^{(s)}=-\frac1s e^{-s\mathcal{P}^{(0)}}\,.
\end{equation}
$\mathcal{P}^{(s)}, \Phi_0$ provides us with another basis in which the charges are integrable. In section \ref{sec:changebasis} we discuss that there are in fact infinitely many such integrable basis, with different charge algebras.

%%%%%%%%%%%%%%%%%%%%%%%%%%%%%%%%%%%%%%%%%%%%%%%%%%%%%%%%%%%%%%%%%%%
\section{Null Boundary Symmetry (NBS)  Algebra, \emph{3d}  Gravity Case}\label{sec:3dgrav}
%%%%%%%%%%%%%%%%%%%%%%%%%%%%%%%%%%%%%%%%%%%%%%%%%%%%%%%%%%%%%%%%%%%

As the next case, we consider the three-dimensional Einstein gravity described by the action and field equations
\be\label{action-EoM-3d}
S=\frac{1}{16\pi G}\int \textrm{d} ^3 x\ \sqrt{-g}\left( R-2\Lambda\right), \qquad    \mathcal{E}_{\mu \nu}:= R_{\mu \nu} - 2\Lambda g_{\mu \nu}=0.
\ee 
Depending on $\Lambda$, $\Lambda<0, \Lambda=0, \Lambda>0$ we respectively have AdS$_3$, flat or dS$_3$ gravities. All solutions to the respective theories are locally AdS$_3$, flat or dS$_3$. The AdS$_3$ case admits BTZ black hole solutions \cite{Banados:1992gq, Banados:1992wn} once we identify one of the spatial directions on a circle. Similarly in flat space, there are the so-called flat space cosmologies \cite{Cornalba:2002fi, Cornalba:2003kd, Bagchi:2012xr}, that are solutions with a cosmological horizon. The dS$_3$ solution itself also has a cosmological horizon, e.g. see \cite{Spradlin:2001pw}.  We will adopt the  $r, v, \phi$ coordinates and take $\phi$ to be $2\pi$ periodic, $\phi\simeq \phi+2\pi$. Since we will be focusing on {the} behaviour of solutions near a null surface, the value of $\Lambda$ will not be of relevance and we leave it free.

{A general family of solutions to the AdS$_3$ theory with $r=0$ as a null surface is specified by three functions of $v,\phi$, see appendix \ref{appen:3d-solutions}. This constitutes the maximal configuration of the $3d$ phase space as it is labelled by three codimension one functions.  We do not use this family of solutions for our charge analysis because it is written in a coordinate system which makes the charge computations rather cumbersome.  However, we still present this solution to convey the idea that the three-functions family of near null surface geometry constructed below, can be extended to a full solution away from $r=0$ surface. }

Let us consider a codimension one null hypersurface $\mathcal{N}$ at $r=0$ and  adopt a Gaussian-null-like coordinate system as follows. Let $v$ be the `advanced time' coordinate along $\mathcal{N}$ such that a null surface is defined by
\begin{equation}
    g^{\mu \nu} \partial_\mu v \partial_\nu v =0\, .
\end{equation}
A ray can be defined as the vector tangent to this surface, $k^\mu = \eta\, g^{\mu \nu} \partial_\nu v$, where $\eta$ is an arbitrary non-zero function. Let $r$ be the affine parameter of the generator $k^\mu$ such that $k^{\mu} = \frac{\d x^{\mu}}{\d r}=\delta^\mu_r$. The last coordinate $\phi \sim \phi + 2 \pi$ is chosen as a parameter constant along each ray, $k^\mu \partial_\mu \phi =0$. With this choice of coordinates metric components are restricted as,
\begin{equation}
    g^{vv}=0 \, , \qquad g^{v r}=\frac{1}{\eta} \, , \qquad g^{v \phi} =0\, ,
\end{equation}
or
\begin{equation}
    g_{ r r}=0 \, , \qquad g_{v r}=\eta \, , \qquad g_{ r \phi} =0\, ,
\end{equation}
where $\eta = \eta(v, \phi)$.\footnote{$k\cdot \nabla k =0$ leads to $\partial_r \eta =0$. {One can perform the analysis without this condition, considering  
$r$-dependent $\eta$. In this case new functions will appear in the symmetry generators which are trivial in the sense that there are no surface charges associated with them. One can use these trivial diffeomorphisms to set sub-leading terms in $g_{vr}$ component of metric, or equivalently $\eta$, and then we are left with only $r$ independent $\eta$. This point, in a very similar $4d$ setup, was discussed in some detail in \cite{Adami:2020amw}. Since the analysis is essentially the same we do not repeat them here.}} Therefore, the most general form of the line-element in the coordinate system constructed above is of the form
\begin{equation}\label{3d-NH-metric}
    \d s^2= - F \d v^2 + 2 \,\eta \d v \d r +2\, f\, \d v \d \phi +h \d \phi^2 \,,
\end{equation}
where $F$, $f$ and $h$ are some functions of $x^{\mu}$. We will be interested in near null surface $r=0$ expansion of the above general metric, for which 
\begin{subequations}\label{r-expansion-3d}
    \begin{align}
       F(v,r,\phi) =& F_0(v,\phi) + r F_1(v,\phi) +\mathcal{O}(r^2) \\
       f(v,r,\phi) =& f_0(v,\phi) + r f_1(v,\phi) +\mathcal{O}(r^2) \\
       h(v,r,\phi) =& \Omega(v,\phi)^2 + r h_1(v,\phi) +\mathcal{O}(r^2)
    \end{align}
\end{subequations}
Since $r=0$ is a null hypersurface, we must have $g^{rr}|_{r=0}=0$. Hence,
\begin{equation}\label{Null-condition}
    F_0= - \left(\frac{f_0}{\Omega} \right)^2 \, .
\end{equation}

The near $r=0$ metric has seven functions of $v,\phi$, which are $\eta, h_0, h_1, f_0, f_1, F_0, F_1$. Besides \eqref{Null-condition} there are three other relations among them resulting from the EoM \eqref{action-EoM-3d} which may be imposed order by order in $r$. At  zeroth order $\mathcal{E}_{v v}=0$, $\mathcal{E}_{v \phi}=0$ and $\mathcal{E}_{\phi \phi}=0$ yield
\begin{equation}\label{v-v}
\partial_v^2 \Omega + \frac{1}{2}\left(\Gamma-\frac{\partial_{v}\eta}{\eta} + \frac{f_0 \partial_\phi \eta }{ \Omega^2 \eta}  \right) \chi -  \partial_\phi \left( \frac{\frac{1}{2}\partial_\phi F_0+\partial_v f_0}{\Omega}\right)  =0
\end{equation}
\begin{equation}\label{v-phi}
\partial_v\Upsilon + \Omega \partial_\phi \Gamma + \frac{\chi\,\partial_\phi \eta }{\eta} +f_0 \partial_\phi \left( \frac{\Upsilon}{\Omega^2}\right)-2 \Omega \partial_{\phi} \left( \frac{f_0 \Upsilon}{\Omega^3}\right) =0
\end{equation}
\begin{equation}\label{phi-phi}
    \partial_{\phi}{\left(\frac{f_{1}-\partial_{\phi}\eta}{\Omega}\right)}-\partial_{v}{\left(\frac{h_{1}}{\Omega}\right)}+\frac{(\partial_{\phi}{\eta})^2}{2\eta\Omega}-\Lambda{\eta\Omega}-\frac{f_{1}^2+F_{1}h_{1}}{2\eta\Omega}=0
\end{equation}
with
\begin{subequations}
    \begin{align}
        \Upsilon:=&\frac{ f_{0}h_{1}}{\eta\Omega}-\frac{f_{1}\Omega}{\eta} \, \label{Upsilon-3d} \\
    \Gamma :=& -\frac{F_{1}}{\eta}-\frac{\partial_{v}\eta}{\eta}- \frac{h_1 f_0^2}{\eta \Omega^4}+\frac{f_0 \partial_\phi \eta}{\eta \Omega^2}+ \frac{2f_0 \Upsilon}{\Omega^3} \label{Gamma-3d}\\
    \chi:=&\partial_v \Omega - \partial_\phi \left( \frac{f_0}{\Omega}\right)
    \end{align}
\end{subequations}
The rest of equations of motion determine higher orders in metric expansion in terms of the lower ones. {Eq.\eqref{v-v} may be viewed as an algebraic equation for $\Gamma$ and \eqref{v-phi}, \eqref{phi-phi} as first order differential equations for $\Upsilon$ and $f_1$.\footnote{The $v$ dependent integration functions will not play any role in our charge analysis and hence we do not consider them.} }

%%%%%%%%%%%%%%%%%%%%%%%%%%%%%%%%%%%%%%%%%%%%%%%%%%%
\subsection{NBS generating  vector fields}
%%%%%%%%%%%%%%%%%%%%%%%%%%%%%%%%%%%%%%%%%%%%%%%%%%%
The vector field
\begin{equation}\label{KV-3D}
    \begin{split}
        \xi^v =& T\\
        \xi^r =& r (\partial_v T- W) + \frac{r^2 \partial_\phi T}{2 \Omega^2}  \left( f_1 + \partial_\phi \eta -\frac{f_0 h_1}{\Omega^2}\right) +\mathcal{O}(r^3)\\
        \xi^\phi =&  Y - \frac{r\eta \partial_\phi T}{\Omega^2}+ \frac{r^2\eta h_1\partial_\phi T}{2 \Omega^4} +\mathcal{O}(r^3)
    \end{split}
\end{equation}
preserves the line-element \eqref{3d-NH-metric}, where $T$, $Y$ and $W$ are some functions of $v$ and $\phi$. The symmetry generating  vector field \eqref{KV-3D} preserves location of the null surface $r=0$, which we take to be the boundary of our spacetime. To see this explicitly, we note that 
\begin{equation}
\begin{split}
\mathcal{L}_{\xi}g^{rr}=&\xi^{\mu}\partial_{\mu}g^{rr}-2g^{r\mu}\partial_{\mu}\xi^{r}\\
=&\xi^{r}\partial_{r}g^{rr}+\xi^{v}\partial_{v}g^{rr}+\xi^{\phi}\partial_{\phi}g^{rr}-2g^{rr}\partial_{r}\xi^{r}-2g^{rv}\partial_{v}\xi^{r}-2g^{r\phi}\partial_{\phi}\xi^{r}\nonumber
\end{split}
\end{equation}
therefore, $\delta_\xi g^{rr}|_{r=0}=0$. 
One may  compute the algebra of NBS generating vector fields \eqref{KV-3D} using the adjusted bracket, yielding
\begin{equation}\label{3d-NBS-KV-algebra}
    [\xi( W_1, T_1, Y_1), \xi( W_2, T_2, Y_2)]_{_{\text{adj. bracket}}}=\xi( W_{12}, T_{12}, Y_{12})
\end{equation}
where 
\begin{subequations}\label{W12-T12-Y12}
\begin{align}
    &  T_{12}=T_1 \partial_v T_2 - T_2 \partial_v T_1 +Y_1 \partial_\phi T_2 - Y_2 \partial_\phi T_1\\
    &  W_{12}=   T_1 \partial_v W_2 - T_2 \partial_v W_1 +Y_1\partial_\phi  W_2- Y_2\partial_\phi  W_1+ \partial_v Y_1 \partial_\phi T_2 -\partial_v Y_2 \partial_\phi T_1 
    \\
&  Y_{12}= Y_1\partial_\phi  Y_2- Y_2\partial_\phi  Y_1+T_1 \partial_v Y_2 - T_2 \partial_v Y_1 \, ,
\end{align}
\end{subequations}
This is Diff$(C_2)\oplus$ Weyl$(C_2)$, where $C_2$ stands for the null cylinder parameterised by $v,\phi$ and Weyl$(C_2)$
is the Weyl scaling on this cylinder; $T, Y$ are generators of  Diff$(C_2)$ and $W$ that of Weyl$(C_2)$.\footnote{As our analysis here indicates, if we repeat the null boundary symmetry analysis for general dimension $d$, we will find Diff$(C_{d-1})\oplus$ Weyl$(C_{d-1})$, where $C_{d-1}$ is the $d-1$ dimensional null ``cylinder'' one would find at $r=0$ and Weyl$(C_{d-1})$ denotes Weyl scaling on $C_{d-1}$. It is clear that generators of this algebra consist of $d=(d-1)+1$ functions over $C_{d-1}$.}

Under transformations generated by these vector fields,  fields transform as
\begin{subequations}
    \begin{align}
        \delta_\xi \eta =& T \partial_v \eta + 2 \eta \partial_v T - \eta W +Y \partial_\phi \eta - \frac{f_0 \eta }{\Omega^2} \partial_\phi T\\
              \delta_\xi \Omega = & T \partial_v \Omega + \partial_\phi \left( Y \Omega\right)+\frac{f_0 }{\Omega}  \partial_\phi T \\
              \delta_\xi f_0 = & \partial_v \left( T f_0\right) +\partial_\phi \left( Y f_0\right)+ \Omega^2 \partial_v Y -F_0 \partial_\phi T 
     \end{align}
\end{subequations} 
\begin{subequations}
\begin{align}
\hspace*{-5mm}\delta_{\xi}{\Upsilon}&=T\partial_{v}\Upsilon+ Y\partial_{\phi}\Upsilon +2\Upsilon\partial_{\phi}Y +\Omega\partial_{\phi}\left( W -\frac{f_{0}\partial_{\phi}T}{\Omega^2}\right)-\Omega \left(\Gamma -\frac{2f_{0}\Upsilon}{\Omega^3}+\frac{2\chi}{\Omega}\right)\partial_{\phi}T\\
\hspace*{-5mm}\delta_{\xi}\Gamma&=\partial_{v}{(T\Gamma)}+Y\partial_{\phi}\Gamma-\partial_{v}W+\partial_{v}{(\frac{f_{0}\partial_{\phi}T}{\Omega^2})}
+\frac{f_{0}\partial_{\phi}W}{\Omega^2}-\frac{f_{0}}{\Omega^2}\left(\Gamma\partial_{\phi}T+\partial_{\phi}{(\frac{f_{0}\partial_{\phi}T}{\Omega^2})}\right)\\
\hspace*{-5mm}\delta_\xi \chi&= \partial_v (T \chi) +\partial_\phi (Y \chi) \,. 
 \end{align}
\end{subequations}
%%%%%%%%%%%%%%%%%%%%%%%%%%%%%%%%%%%%%%%%%%%%%%%%%%%%%%%%%%%%%%%%%%%%%%%%%
\subsection{Surface charges}
%%%%%%%%%%%%%%%%%%%%%%%%%%%%%%%%%%%%%%%%%%%%%%%%%%%%%%%%%%%%%%%%%%%%%%%%%

The Iyer-Wald surface charge \cite{Iyer:1994ys} is given by
\begin{equation}\label{3dnonintch}
        \slashed{\delta} Q_{\xi} := \oint_{\partial \Sigma} \mathcal{Q}^{\mu \nu}_\xi[g ; \delta g] \d x_{\mu \nu}
\end{equation}
with
\begin{equation}\label{charge-var-non-int}
    \mathcal{Q}^{\mu \nu}_\xi =\frac{\sqrt{-g}}{8 \pi G}\, \Big( h^{\lambda [ \mu} \nabla _{\lambda} \xi^{\nu]} - \xi^{\lambda} \nabla^{[\mu} h^{\nu]}_{\lambda} - \frac{1}{2} h \nabla ^{[\mu} \xi^{\nu]} + \xi^{[\mu} \nabla _{\lambda} h^{\nu] \lambda} - \xi^{[\mu} \nabla^{\nu]}h \Big),
\end{equation}
where $h_{\mu \nu}= \delta g_{\mu \nu}$  is a metric perturbation and $h= g^{\mu \nu}h_{\mu \nu}$. We take $\Sigma$ to be a constant $v$ hypersurface and $\partial \Sigma$ to be its cross-section with {the hypersurface} $r=0$. The surface charge on the given co-dimension two surface $\partial \Sigma$ is then
\begin{equation}\label{charge-variation''}
        \slashed{\delta} Q_{\xi} = \frac{1}{16 \pi G}\int_{0}^{2 \pi} \d \phi \left[ W \delta \Omega + Y \delta \Upsilon + T \slashed{\delta} \mathcal{A}\right] \, , 
\end{equation}
with
\begin{equation}
       \slashed{\delta} \mathcal{A}=-2 \delta \chi - \Gamma \delta \Omega + \frac{ f_0}{\Omega^2} \delta\Upsilon {+} \partial_\phi \left( \frac{ f_0 \delta \Omega}{\Omega^2}\right) + \frac{ \chi\delta \eta}{\eta} 
\end{equation}
As we see in the basis \eqref{KV-3D} the charges are not integrable for state independent $W,Y,T$. One may use {the} modified bracket method to separate its flux and integrable parts, as done in Appendix \ref{BT-MB-Appendix}. However, in what follows we show  there exists a basis in which the charges are integrable. 

%%%%%%%%%%%%%%%%%%%%%%%%%%%%%%%%%%%%%%%%%%%%%%%%%%%%%%%%%%%%%%%%%%%%%%%%%
\subsection{Integrable basis for the charges}\label{sec:integrabe-3d-basis}
%%%%%%%%%%%%%%%%%%%%%%%%%%%%%%%%%%%%%%%%%%%%%%%%%%%%%%%%%%%%%%%%%%%%%%%%%
Let us consider the  state/field-dependent transformations,
\begin{equation}\label{hat-generators-3d}
    \hat{W}=W- T \Gamma  {-}\ \frac{f_0}{\Omega^2} \ \partial_\phi T \, , 
    \qquad \hat{Y}=Y+T \frac{f_0 }{\Omega^2} \, , \qquad \hat{T} = \frac{\chi}{\Xi^{(s)}} T \, ,
\end{equation}
or conversely, 
\begin{equation}\label{Int-basis-generators}
\begin{split}
T=\frac{\Xi^{(s)}}{\chi}\hat{T}, \hspace{1 cm} Y=\hat{Y}-\frac{f_{0}\Xi^{(s)}}{\chi\Omega^2}\hat{T},\hspace{1 cm}
W=\hat{W}+\frac{\Gamma\Xi^{(s)}}{\chi}\hat{T}{+}\frac{f_{0}}{\Omega^2}\partial_{\phi}\left(\frac{\Xi^{(s)}\hat{T}}{\chi}\right)\ ,
\end{split}
\end{equation}
where $s$ is an arbitrary real number and
\begin{equation}\label{Def-Xi-01}
      \Xi^{(s)} := \left(\frac{\chi^2}{\eta} \right)^{s}\ .
\end{equation}
In this basis the charge variation takes the form 
\begin{equation}\label{3d-integrable-charge-variation}
    \delta Q_{\xi} = \frac{1}{16 \pi G}\int_{0}^{2 \pi} \d \phi\, \left( \hat{W} \delta \Omega + \hat{Y} \delta \Upsilon +\hat{T} \delta \mathcal{P}^{(s)}\right)
\end{equation}
with
\begin{equation}\label{charges-3d}
 \mathcal{P}^{(s)}=
  \begin{cases} 
   - \frac{1}{s}\,\Xi^{(s)} & \text{if } s \neq 0 \\
   -\ln{\Xi^{(1)}}       & \text{if } s =0
  \end{cases}
\end{equation}
which is clearly integrable if $\hat W, \hat Y, \hat T$ are taken to be field-independent, i.e. $\delta\hat W=\delta\hat Y=\delta\hat T=0$. 

We note that the equation of motion for $\mathcal P^{(s)}$ is
\begin{equation}\label{calEs-EoM}
  \begin{cases} 
  \frac{\partial_v \mathcal P^{(s)}}{\mathcal P^{(s)}} - \frac{f_0}{\Omega^2}\frac{\partial_\phi \mathcal P^{(s)}}{\mathcal P^{(s)}}+s\Gamma -2s \partial_\phi\left(\frac{f_0}{\Omega^2}\right) &=0,\qquad s\neq 0,\\ & \\
    \partial_v \mathcal{P}^{(0)} -\frac{f_0}{\Omega^2}\partial_\phi \mathcal P^{(0)}-\Gamma +2 \partial_\phi\left(\frac{f_0}{\Omega^2}\right) &=0.
\end{cases}\end{equation}
So, among the six fields $\{\Omega, \, \mathcal{P}^{(s)}, \, \Upsilon; f_0, \, f_1 , \, \Gamma \}$ one can solve \eqref{calEs-EoM}, \eqref{v-phi} and \eqref{phi-phi} to obtain  $f_0, \, f_1 , \, \Gamma$ in terms of $\Omega, \, \mathcal{P}^{(s)}, \, \Upsilon$, up to two $v$-dependent integration constants. Nonetheless,  as one can  see explicitly from \eqref{3d-integrable-charge-variation}, only the three arbitrary fields $\Omega, \, \mathcal{P}^{(s)}, \, \Upsilon$, and not the two $v$-dependent functions,  appear in the surface charge expressions. 
Therefore, the  solution phase space  can be characterized by these three functions on codimension one $r=0$ null hypersurface.

Being field-dependent, one should use the adjusted Lie bracket \cite{Barnich:2010eb, Compere:2015knw} when computing the bracket of two symmetry generators. This  yields
\begin{equation}\label{NHKV-algebra-3d}
    [\xi(\hat W_1, \hat T_1, \hat Y_1), \xi(\hat W_2, \hat T_2, \hat Y_2)]_{_{\text{adj. bracket}}}=\xi(\hat W_{12}, \hat T_{12}, \hat Y_{12})
\end{equation}
where 
\begin{subequations}\label{NHKV-algebra-3d-01}
\begin{align}
    & \hat T_{12}=s(\hat T_1\hat W_2-\hat T_2\hat W_1)+\hat Y_1\partial_\phi \hat T_2-\hat Y_2\partial_\phi \hat T_1+ (2s-1)(\hat T_1 \partial_\phi \hat Y_2 - \hat T_2 \partial_\phi \hat Y_1 )\\
    & \hat W_{12}= \hat{Y}_1\partial_\phi \hat W_2 -\hat{Y}_2\partial_\phi \hat W_1   
    \\
& \hat Y_{12}=\hat Y_1\partial_\phi \hat Y_2-\hat Y_2\partial_\phi \hat Y_1,
\end{align}
\end{subequations}
for both generic $s$ and $s=0$. As we see the above algebra does not involve derivatives w.r.t. $v$ parameter. 

To read off the algebra of charges $\Omega, \Upsilon, \mathcal{P}^{(s)}$ we need to  know the variations of charges under diffeomorphisms, which upon equations of motion \eqref{v-v} and \eqref{v-phi}, they can be simplified to
\begin{subequations}
    \begin{align}
        \delta_\xi \Omega= &  -s\mathcal{P}^{(s)} \hat{T} + \partial_\phi ( \hat{Y} \Omega )\, , \\
        \delta_\xi \mathcal{P}^{(s)}=  & s \hat{W} \, \mathcal{P}^{(s)} +2s \,\mathcal{P}^{(s)} \,\partial_\phi \hat{Y} + \partial_\phi \mathcal{P}^{(s)} \hat{Y} \, ,\\
        \delta_{\xi}{\Upsilon}=& (2s-1)\,  \partial_\phi \mathcal{P}^{(s)} \hat{T} +2s \mathcal{P}^{(s)} \partial_\phi \hat{T} +\hat{Y} \partial_\phi \Upsilon +2 \Upsilon \partial_\phi \hat{Y}+ \Omega \partial_\phi \hat{W} \, ,
    \end{align}
\end{subequations}
for $s \neq0$ and
\begin{subequations}
    \begin{align}
       \delta_\xi \Omega= &  \hat{T} + \partial_\phi ( \hat{Y} \Omega )\, , \\
      \delta_\xi \mathcal{P}^{(0)}=& -\hat{W} -2\partial_\phi \hat{Y}+\partial_\phi \mathcal{P}^{(0)} \hat{Y}    \\
      \delta_{\xi}{\Upsilon}=& -  \hat{T} \partial_\phi \mathcal{P}^{(0)}- 2 \partial_\phi \hat{T} +\hat{Y} \partial_\phi \Upsilon +2 \Upsilon \partial_\phi \hat{Y}+ \Omega \partial_\phi \hat{W} \, ,
   \end{align}
\end{subequations}
for $s=0$.

The algebra of charges in new basis is hence
 \begin{equation}\label{Charge-algebra}
 \begin{split}
  \hspace*{-1cm}    \{ Q^{(s)}(\xi_1),Q^{(s)}(\xi_2)\}&= Q^{(s)}(\xi_{12}) \\
      &+ \frac{1}{16 \pi G}\, \delta_{s,0} \int_{0}^{2\pi} \d \phi \left[  (\hat{W}_1 \hat{T}_2 - \hat{W}_2 \hat{T}_1 ) -2 ( \hat{Y}_1 \partial_\phi \hat{T}_2 - \hat{Y}_2 \partial_\phi \hat{T}_1) \right]\, ,
 \end{split}
 \end{equation}
which is the algebra of symmetry generators up to the central terms appearing in {the} $s=0$ case. {The charges, $\Omega,\mathcal P^{(s)},\Upsilon$, depend on the coordinates $(v,\phi)$ and their  algebra is
\begin{subequations}\label{3d-charge-algebra}
    \begin{align}
        & \{ \Omega(v,\phi), \Omega (v,\phi') \} =0 \, ,\label{3d,a'}\\
        & \{\mathcal{P}^{(r)}(v,\phi), \mathcal{P}^{(s)}(v,\phi') \} =0 \, , \label{3d,b'}\\
        & \{\Omega(v,\phi), \mathcal{P}^{(s)}(v,\phi') \} = 16\pi G \left( - s \mathcal{P}^{(s)}(v,\phi)+ \delta_{s,0}\right) \, \delta(\phi-\phi') \, , \label{3d,c'}\\
        & \{\Upsilon(v,\phi), \Upsilon(v,\phi') \} = 16\pi G\left(\Upsilon(v,\phi') \partial_\phi -\Upsilon(v,\phi) \partial_{\phi'} \right) \delta(\phi-\phi') \, , \label{3d,d'}\\
        & \{\Upsilon(v,\phi), \Omega(v,\phi') \}= -16 \pi G\, \Omega(v,\phi) \,\partial_{\phi'}\, \delta(\phi-\phi') \, ,\label{3d,e'}\\
        & \{\Upsilon(v,\phi), \mathcal{P}^{(s)}(v,\phi') \}=  16\pi G\left( - \mathcal{P}^{(s)}(v,\phi)\partial_{\phi'} + (2s-1) \, \mathcal{P}^{(s)}(v,\phi')\,\partial_\phi +2\delta_{s,0}\partial_{\phi'}\right) \delta(\phi-\phi') \, .\label{3d,f'}
    \end{align}
\end{subequations}}
If we Fourier expand the charges
\begin{equation}
\begin{split}
    \mathcal{P}^{(s)}(v,\phi) :=8 G\,\sum_n \mathcal{P}^{(s)}_n(v) e^{{-}in\phi}&,\qquad
    \Omega(v,\phi) :=8 G\,\sum_n \Omega_n(v) e^{{-}in\phi},\\
    \Upsilon(v,\phi) :=&8 G\,\sum_n \Upsilon_n(v) e^{{-}in\phi},
\end{split}
\end{equation}
and upon quantisation $i\{,\}\to [ ,]$, we have
\begin{subequations}\label{3d-charge-algebra''}
    \begin{align}
        & [\Omega_m(v), \Omega_n(v) ] =0 \, , \\
        & [\mathcal{P}^{(r)}_{m}(v), \mathcal{P}^{(s)}_{n}(v) ] =0 \, , \\
        & [\Omega_{m}(v), \mathcal{P}^{(s)}_{n}(v) ] =-i\,s\, \mathcal{P}^{(s)}_{m+n}(v)+ \frac{i}{8 G}\, \delta_{s,0} \,\delta_{m+n,0}  \, , \\
        & [\Upsilon_{m}(v), \Upsilon_{n}(v) ] =(m-n)\, \Upsilon_{m+n}(v)\, \, , \\
        & [\Upsilon_{m}(v), \Omega_{n}(v) ]= -n \, \Omega_{m+n}(v)\,  ,\\
        & [\Upsilon_{m}(v), \mathcal{P}^{(s)}_{n}(v)]= \left((2s-1)\, m -n\right)\, \mathcal{P}^{(s)}_{m+n}(v)+ \frac{n}{4 G} \,\delta_{s,0} \,\delta_{m+n,0}  \, .\label{Y-E-algebra}
    \end{align}
\end{subequations}
While the charges are in general $v$ dependent their algebra is not; meaning that the algebra takes the same form for all $v$. 

We close this part by some comments:
\begin{itemize}

\item The algebra \eqref{3d-charge-algebra''} consists of a Witt algebra, spanned by $Y$, semidirect sum with the ``supertranslation'' $T$ (of conformal weight $2s$) and the current $W$ (of conformal weight 1). The two supertranslations commute among each other for $s=0$.
    \item The form of metric near $r=0$ null surface \eqref{3d-NH-metric} and \eqref{r-expansion-3d}, may be viewed as near horizon metric in which $\Omega$, which is the area density of the bifurcation surface (circle in the $3d$ case), is the ``entropy density''. Note that $\Omega$ is  the charge associated with $\hat{W}$ \eqref{3d-integrable-charge-variation} and in our original geometry  involves scaling in $r$ direction accompanied by a field-dependent diffeomorphism in $v$ direction, \emph{cf.} \eqref{KV-3D}, \eqref{hat-generators-3d}.  
    \item Our results for the NBS generating  vectors \eqref{hat-generators-3d} and the associated charges $\Omega, \Upsilon, \mathcal{P}^{(s)}$ and their algebra \eqref{charges-3d} are all independent of the cosmological constant. $\Lambda$ appears only in one of our  equations of motion \eqref{phi-phi}, and does not enter in our charge analysis in any crucial way. Hence,  our results for charges and symmetries are  valid for AdS$_3$, $3d$ flat space and dS$_3$ cases.

    This is physically expected because we are considering  an expansion around {the} null surface $r=0$, and  our analysis is local and independent of the asymptotic and global properties of the solution.
    \item There is an interesting sublagebra of \eqref{3d-charge-algebra} spanned by $\Omega_m$ and $\Upsilon_m$ which is a semidirect sum of the Witt algebra generated by $\Upsilon_m$ with an Abelian current $\Omega_m$. In the terminology used in \cite{Parsa:2018kys}, it is $W(0,0)$ algebra. This algebra was found as near horizon symmetry algebra of $3d$ black holes \cite{Donnay:2015abr} (see also \cite{Grumiller:2019fmp}). But one should note that in our case the charges have arbitrary $v$ dependence, while that of \cite{Donnay:2015abr, Grumiller:2019fmp} is $v$ independent. Moreover, here $\Omega_m$ is associated with the Weyl scaling  on the null surface $r=0$ but instead in \cite{Donnay:2015abr} it was associated with supertranslation on the given null surface. However, in both cases, the zero-mode charges $\Omega_0$ and $\Upsilon_0$ are respectively proportional to {the} entropy and angular momentum of the black holes whose horizon is taken to be the null surface we are expand about.
    \item The algebra \eqref{3d-charge-algebra} has another sublagebra spanned by $\mathcal{P}^{(s)}_{m}(v)$ and $\Upsilon_{m}(v)$. Modulo the arbitrary $v$ dependence of the charges, it coincides with $W(0,-2s-1)$ \cite{gao2011low,Parsa:2018kys}. This algebra is an algebraic deformation of BMS$_3$ \cite{Parsa:2018kys} where $\mathcal{P}^{(s)}_{m}(v)$  describes spin-$2s$ supertranslations. 
\item For the  $s=0$ case and as \eqref{3d,a'}, \eqref{3d,b'} and \eqref{3d,c'} show, $\Omega(v,\phi)$ and $\mathcal{P}^{(0)}(v,\phi)$ form a Heisenberg algebra and one can treat $\mathcal{P}^{(0)}(v,\phi)$  as conjugate momentum of the field $\Omega(v,\phi)$; these commutation relations upon the usual replacement of Poisson brackets by commutators, $i\,\{\,, \, \} \to [\, , \,]$,
may be viewed  as the ``equal time'' (equal $v$) canonical commutation relations of a two-dimensional field theory defined on {the $r=0$ surface}.
\item The algebra for generic $s$ case can be constructed using the $s=0$ algebra through
\begin{equation}\label{G-s-01}
    \mathcal{P}^{(s)}= -\frac{1}{s} \exp{\left(-s \,\mathcal{P}^{(0)}\right)}\,.
 \end{equation}
   \item The structure constants of the algebra \eqref{3d-charge-algebra''} and its $2d$ counterpart \eqref{2d-Phi-Xi-algebra} are both $v$ independent, while the charges are $v$ dependent. We expect this to be the case for all integrable Lie-algebras related to these algebras by a change of basis (see discussions of the next section). One should, however, note that this is not the case for the algebra of the original symmetry generators (before coming to the integrable basis), \eqref{3d-NBS-KV-algebra} and \eqref{W12-T12-Y12} for $3d$ case and \eqref{2d-algebra}, for the $2d$ case; they involve derivatives w.r.t $v$. Therefore,  the algebra of integrable part of charges obtained through the modified bracket method  (see appendix \ref{BT-MB-Appendix}) would also have $v$ derivatives. 
   \end{itemize}

%%%%%%%%%%%%%%%%%%%%%%%%%%%%%%%%%%%%%%%%%%%%%%%%%%%
\section{Change of Basis for Integrable Charges }\label{sec:changebasis}
%%%%%%%%%%%%%%%%%%%%%%%%%%%%%%%%%%%%%%%%%%%%%%%%%%%

In the previous sections, we discussed a specific {field-dependent} change of basis {that} rendered non-integrable charges integrable, yielding Heisenberg-type algebra among the integrable charges {in} the new basis. {Given} this specific $2d, 3d$ examples one is led to two questions:
\begin{enumerate}
    \item Is it possible to always make a set of given non-integrable charges integrable by a {field-dependent} change of basis? If not, when does the integrable basis exist?
    \item Once in an integrable basis, can we still make {a} further change of basis and remain with integrable charge? 
\end{enumerate}

In this work, we mainly restrict ourselves to the $2d, 3d$ setups we have explored in the previous two sections for which (by construction) the answer to the first question is affirmative. While the full analysis of the first question is postponed to an upcoming publication \cite{progress-2}, we will briefly discuss it in the last section \ref{sec:conclusion}. In this section, we  focus on the second question. Assuming we have a basis in which the charges are integrable, in subsection \ref{sec:4.1} we present a general formulation of change of basis among integrable charges. Then in the next two subsections, we apply this general formulation to $2d$ and $3d$ cases.

\subsection{Change of basis, general formulation}\label{sec:4.1}

Suppose that we have a set of integrable surface charges generated by {the} symmetry generators $\mu^i$, 
\begin{equation}\label{charge-variation-generic}
    \delta Q=\int  \mu^i \delta Q_i\, \d{}^{d-2}x\, , \qquad i=1,2,\cdots, N.
\end{equation}
Integrability of $Q_i$ means that $\mu^i$ are field-independent functions,\footnote{As our $2d, 3d$ examples show (see also \cite{Adami:2020amw} for 
$4d$ examples), while the integral \eqref{charge-variation-generic} is over a codimension two surface, constant $v$ slices on $r=0$ in our setup, $\mu^i$ are in general functions over a codimension one surface.} that is $\delta \mu^i / \delta Q^j=0$. We adopt  a thermodynamical terminology in which $\mu^i$ are referred to as  chemical potentials. Thus the charge variation is integrable on phase space,
\begin{equation}
    Q=\int  \mu^i  Q_i\, \d{}^{d-2}x \, .
\end{equation}

Let us now consider a generic change of basis of the form
\begin{equation}\label{change-of-basis}
    \tilde Q_i=\tilde Q_i[Q_j, \partial^n Q_k] \, ,
\end{equation}
which is a functional of $Q_{i}$ and their derivatives. {The functional $\tilde Q_i$ may be restricted by some physical conditions like being well-defined and real, e.g. excluding  $\tilde Q_i=\frac1{Q_i}$ or $\tilde Q_i=\sqrt{Q_i}$ when charges can be zero or negative, but is otherwise generic. It even need not be one-to-one and may reduce number of charges, restricting us to a subspace in the space of solutions over which the charges are defined. However, here for simplicity we assume $ \tilde Q_i$ to be one-to-one. Of course  one can make further restrictions to desired subspaces after the change of basis.} 

We can now impose the condition that the charges are integrable in the new basis, as we have already implicitly assumed in \eqref{change-of-basis}. In the new basis, the charge variation can be written as
\begin{equation}\label{charge-variation-new}
    \delta Q=\int  \tilde{\mu}^i \delta \tilde{Q}_i\, \d{}^{d-2}x\, ,
\end{equation}
where $\tilde\mu^i$ are the chemical potentials in the new basis and $\delta \tilde{\mu}^i / \delta \tilde{Q}_j=0$. Equating  \eqref{charge-variation-generic} and \eqref{charge-variation-new}, one can read the relation between the chemical potentials in the two basis,
\begin{equation}\label{chemical-pot-mapping}
    \tilde \mu^i=\frac{\delta Q_j}{\delta \tilde Q_i}\ \mu^j.
\end{equation}
where $\frac{\delta Q_j}{\delta \tilde Q_i}$ denotes {the} variation of $Q_i$ w.r.t $\tilde Q_j$ and we assume this variation exists and is well-defined. We will show below that $\frac{\delta Q_j}{\delta \tilde Q_i}$ can in general be an operator involving spatial  derivatives along the $d-2$ directions, acting on $\mu^j$. In essence, one may start with an arbitrary $\tilde Q_i=\tilde Q_i[Q_j]$ and integrability condition fixes $\mu^i$ in terms of $\tilde{\mu}^i$.
As we see the two chemical potentials are related linearly by field-dependent coefficients. Note also that the charge variation $\delta Q$ is the same in the two bases (imposed through our integrability requirement). The integrated charge $Q$ is however \emph{not} necessarily the same in the two bases;  they are only equal iff  $\tilde Q_i$ are linear combinations of $Q_i$.

The above conditions are a bit abstract and formal. One can explore them further introducing,
\begin{equation}\label{Ch-den-var}
    \delta \tilde Q_i := \Pi_{i}{}^{j} \delta  Q_j \, ,
\end{equation}
with
\begin{equation}
    \Pi_{i}{}^{j}= \sum_{p=0} (\Pi_{i}{}^{j}){}^{A_1\cdots A_p}  [Q_i] \, \partial_{A_1\cdots A_p} \, , \qquad (\Pi_{i}{}^{j}){}^{A_1\cdots A_p} [Q_i]:= \frac{\partial \tilde{Q}_i}{\partial (\partial_{A_1\cdots A_p} Q_j)} \, ,
\end{equation} 
where $A_l=1,\cdots, d-2$ are running over the directions along the codimension two spacelike surface the charges are integrated over, in our case constant $v$ slices on the $r=0$ null surface.\footnote{Note that the charges $Q_i$ and $(\Pi_{i}{}^{j}){}^{A_1\cdots A_p}$ can have general $v$ dependence.}
Substituting \eqref{Ch-den-var} into \eqref{charge-variation-new} and integrating by-part, one can make \eqref{chemical-pot-mapping} more explicit:
\begin{equation}\label{mu-tilde-mu}
\mu^i = \sum_{p=0} (-1)^p \partial_{A_1\cdots A_p} \left[ (\Pi_{j}{}^{i}){}^{A_1\cdots A_p} \tilde{\mu}^j\right].
\end{equation}

\paragraph{Algebra of charges under change of basis.} Let us suppose that the Poisson bracket of the charges in the original basis is of the form of a Lie-algebra
\begin{equation}\label{Q-Lie-algebra}
    \{ Q_i (v,{x}^A), Q_j(v,{y}^B)\}= f_{ij}{}^k({x}^A, {y}^B;v) Q_k (v,{x}^C) -f_{ji}{}^k({y}^B, {x}^A;v) Q_k (v,{y}^C)+ c_{ij}({x}^A, {y}^B;v),
\end{equation}
where $c_{ji}({y}^B, {x}^A;v)=-c_{ij}({x}^A, {y}^B;v)$ and $f_{ij}{}^k$ are $Q_i$ independent while both can have $v,x^A$ {dependence} and can involve derivatives w.r.t. $x^A$ or $y^B$. To ease the notation, whenever there is no confusion we drop $v, x^A$ or the derivative dependence. We assume the charge bracket \eqref{Q-Lie-algebra} satisfies the Jacobi identity, $\{ \{ Q_i, Q_j\}, Q_k\}+\text{cyclic permutation}=0$. After a change of basis to $\tilde Q_i$, the algebra of new charges is hence
\begin{equation}\label{tilde-Q-algebra}
    \{ \tilde Q_i, \tilde Q_j\}=  \Pi_i{}^k\Pi_j{}^l\ \{ Q_k, Q_l\}
\end{equation}
The right-hand-side, as we see is not necessarily of the form $\tilde f_{ij}{}^k \tilde Q_k+ c_{ij}$ for some $\tilde Q$ independent structure constants $\tilde f_{ij}{}^k$ and central charges $c_{ij}$. Remember that $\Pi_i{}^j$ are in general operators involving spacetime derivatives acting on $Q_m$, as well as being matrices. One may ask if $\tilde Q_i$ form an algebra, i.e. if their bracket satisfies Jacobi identity. The answer is not always affirmative because
\begin{equation}\label{Jacobi-tilde-Q}
\begin{split}
   \{ \{ \tilde Q_i, \tilde Q_j\}, \tilde Q_k\}+\text{cyclic perm.} &=  \Pi_i{}^l\Pi_j{}^m \Pi_k{}^n\ \left(\{\{ Q_l, Q_m\}, Q_n\}+\text{cyclic perm.}\right)\\ 
    &+  \Pi_k{}^n \{\Pi_i{}^l\Pi_j{}^m, Q_n\}\{Q_l, Q_m\}+\text{cyclic perm.}
\end{split}
\end{equation}
While the first line in the right-hand-side vanishes the second line does not in general. This is of course expected, as the general change of basis takes us from the algebra to the enveloping algebra of the charges which does not close onto an algebra with the same dimension. Closure of the algebra in the new basis therefore, severely restricts   possible change of basis.

Given the above discussions, three questions are then in order:
\begin{itemize}
    \item[I.] Is it possible to find an integrable basis where {the} charge algebra is a Lie algebra, possibly up to a central extension?
 In the $2d$ and $3d$ examples, we have shown the answer is affirmative: the Heisenberg type algebras are indeed centrally extended Lie-algebras.  We will argue in the next section that in general there always exists an integrable Lie-algebra basis whenever the charges are integrable. 
   \item[II.]  Are there other possible changes of bases that take an integrable charge algebra to other algebras? That is, are there possible $\Pi_i{}^k$ for which the second line of \eqref{Jacobi-tilde-Q} vanishes? 
\item[III.] If there are other integrable  charge algebras, i.e. if the answer to the previous question affirmative, are there Lie-algebras among them? Is the integrable Lie-algebra  basis unique? The answer is no. See for example explicit constructions in \cite{Parsa:2018kys, Safari:2019zmc, Grumiller:2019fmp}. One can argue, based on the experience and analysis of these references, that if the integrable Lie-algebra we start with is rigid/stable then this basis is unique. Otherwise, if the algebra is not stable and admits non-trivial deformations, all the algebraic deformations of the algebra in integrable Lie-algebra basis will also be an integrable Lie-algebra.
\end{itemize}

In  what follows we do not intend to explore the above questions in full generality, while our analysis for the $2d$ case is exhaustive, for the $3d$ case we analyse  some examples of charges and algebras.

%%%%%%%%%%%%%%%%%%%%%%%%%%%%%%%%%%%%%%%%%%%%%%%%%%%%%%%%%%%%%%%%%%%%%%%%%%%%%
\subsection{\emph{2d} gravity -- generic case}
%%%%%%%%%%%%%%%%%%%%%%%%%%%%%%%%%%%%%%%%%%%%%%%%%%%%%%%%%%%%%%%%%%%%%%%%%%%%%

We have already established in section \ref{sec:2dgrav} that there exist some different integrable basis in which the charge algebra is a Lie-algebra. These bases are labeled by a parameter $s$ \eqref{2d-Phi-Xi-algebra}. So, we already see that for this example there exists more than one integrable Lie-algebra basis. One may, however, still ask if our $s$-family of algebras unique or there are other integrable Lie-algebras which could be reached through a general change of basis discussed above. To this end, let us take the $s=0$ case as the starting point. This is especial in our $s$ family because the charge algebra is just {the} Heisenberg algebra 
\begin{equation}\label{2d-Heisenberg}
    \{ Q_i(v), Q_j(v) \}= 16\pi G\,\varepsilon_{ij},\qquad \forall v,
\end{equation}
with charges  $Q_i=\{ \Phi_0, \mathcal{P}^{(0)} \}$ and symmetry generators $\mu^i =\{ \hat{W},\hat{T} \}$.

Suppose that the new basis $\tilde{Q}_i$ are non-singular functions of original ones. Using properties of Poisson brackets, one can write Poisson bracket of charges in new basis as
\begin{equation}\label{new-basis-no-derivative-2d'}
    \{ \tilde{Q}_i(v), \tilde{Q}_j(v) \}= 16\pi G\, \Pi_i{}^{k} \,\varepsilon_{kl} \,\Pi_j{}^{l} \,.
\end{equation}
where
$\Pi_i{}^{j}{} =\frac{\partial \tilde{Q}_i}{\partial  Q_j}$. One can treat $\Pi_i{}^j$ as a $2\times 2$ matrix and hence $\Pi_i{}^{k} \,\varepsilon_{kl} \,\Pi_j{}^{l}= \varepsilon_{ij} \det{(\Pi_k{}^l)}$. Note that in the $2d$ case $\Pi_i{}^j$ do not involve derivatives and just matrices (rather than being differential operator-valued matrices). 

One can readily see that the closure of the algebra \eqref{new-basis-no-derivative-2d'}  (the Jacobi identity) does not impose any condition on $\Pi_i{}^j$. However, {the} condition of having a Lie-algebra restricts it as,
\begin{equation}\label{Pi-01'}
    \det{(\Pi_i{}^j)}= \gamma +\beta^i \tilde{Q}_i\, ,
\end{equation}
where $\gamma$ and $\beta^i$ are some constants. The charge algebra \eqref{new-basis-no-derivative-2d'} with \eqref{Pi-01'} can also be written as
\begin{equation}\label{2d-changed-basis}
    \{ \tilde{Q}_i(v), \tilde{Q}_j(v) \}= 16\pi G\, (\varepsilon_{ij} \gamma  + \tilde{Q}_i \alpha_j-\tilde{Q}_j \alpha_i) \,  ,
\end{equation}
where $\alpha_i=\varepsilon_{ij} \beta^j$. 

The special case of $\alpha_i=0$ corresponds to generic canonical transformations on the two-dimensional phase space of $\Phi_0, \mathcal{P}^{(0)}$. So, let us focus on the less trivial cases where $\alpha_i\neq 0$. Without loss of generality one can always rotate the basis in the $2d$ $\Phi_0, \mathcal{P}^{(0)}$ plane such that
$\alpha_2=0, \alpha_1:=  s\neq 0$. Then, \eqref{2d-changed-basis} takes the form
\begin{equation}\label{2d-changed-basis-2}
   i \{ \tilde{Q}_1(v), \tilde{Q}_2(v) \}= 16\pi G\, {i} ( \gamma  -  s\tilde{Q}_2 ) \,  .
\end{equation}
For $\gamma=0$ this algebra reduces to \eqref{2d-Phi-Xi-algebra}. So, as we see in the $2d$ case our earlier construction already covers the most general integrable basis up to a central $\gamma$ term.  That is, our $s$ family of algebras \eqref{2d-Phi-Xi-algebra}, up to the central term $\gamma$, exhausts the Lie-algebras one can achieve by a change of basis.

%%%%%%%%%%%%%%%%%%%%%%%%%%%%%%%%%%%%%%%%%%%%%%%%%%%%%%%%%%%%%%%%%%%%%%%%%%%%%
\subsection{\emph{3d} gravity -- general discussion and fundamental basis for NBS algebra}\label{Change-basis-3d}
%%%%%%%%%%%%%%%%%%%%%%%%%%%%%%%%%%%%%%%%%%%%%%%%%%%%%%%%%%%%%%%%%%%%%%%%%%%%%

In the 3$d$ gravity in our integrable basis we have three charges $\Phi_0(v,\phi), \mathcal{P}^{(s)} (v,\phi), \Upsilon(v,\phi)$ which may collectively call them $Q_i$ which satisfy the algebra \eqref{3d-charge-algebra}.  The most general change of basis may then be parameterised through $\Pi_{i}{}^{j}$ \eqref{Ch-den-var}, which is a 3$\times$3 matrix operator which admit a  series expansion in derivative w.r.t. $\phi$ direction,
\begin{equation}
    \Pi_{i}{}^{j}= \sum_{p=0} (\Pi_{i}{}^{j})_p [Q_k] \, \partial_\phi^n \, , \qquad (\Pi_{i}{}^{j})_p [Q_k]= \frac{\partial \tilde{Q}_i}{\partial (\partial_\phi^p Q_j)} \, ,
\end{equation}
where $(\Pi_{i}{}^{j})_p [Q_i]$ are generic $v,\phi$ dependent coefficients. Eq.\eqref{mu-tilde-mu} can hence be written as
\begin{equation}
    \mu^i = \sum_{p=0} (- \partial_\phi )^p \left( (\Pi_{j}{}^{i})_p\ \tilde{\mu}^j\right) \, .
\end{equation}
The analysis here is more involved compared to the $2d$ case because besides the matrices we are also dealing with derivatives in $\phi$, moreover, the  structure constants of the $3d$ charge algebra \eqref{3d-charge-algebra} are not as simple as that of a  Heisenberg algebra.

To make the analysis of change of basis tractable, we focus on the simplest member of the one-parameter family of the algebras  \eqref{3d-charge-algebra}, the $s=0$ case. In addition, we make another change of basis to bring the algebra to the form of {the} direct sum of two Lie-algebras, which we dub as the ``fundamental NBS algebra''.

\subsubsection{\emph{3d} NBS algebra in ``fundamental basis''} Let us start with the $s=0$ algebra with generators $\Phi_0, \mathcal{P}^{(0)}, \Upsilon$,
which form a Heisenberg algebra {in} semidirect sum with the Witt algebra. Consider the following change of basis, keep $\Phi_0, \mathcal{P}^{(0)}$ and redefine $\Upsilon$ as
\begin{equation}\label{Bar-Upsilon}
\Upsilon=-2\partial_{\phi}\Omega-\Omega\partial_{\phi}\mathcal{P}^{(0)}+16\pi G\,{\cal S}\, .
\end{equation}
If we view $\Upsilon$ as generators of ``superrotations'' (as we have in BMS$_3$ algebra), one can decompose it into ``orbital'' part (the external part) and the ``spin'' part ${\cal S}$ (the internal part). To keep the basis integrable, one should then transform the symmetry generators, {i.e.} the chemical potentials, as
\begin{equation}
    \begin{split}
      \hat{W}= & \tilde{W} -2 \partial_\phi \tilde{Y} +\tilde{Y} \partial_\phi \mathcal{P}^{(0)} \, , \\
        \hat{Y}= & \tilde{Y}\, ,\\
        \hat{T}= & \tilde{T} - \partial_\phi (\Omega \tilde{Y})\, .
    \end{split}
\end{equation}
and assume $\tilde{W}, \tilde{Y}, \tilde{T}$ to be independent of charges in the new basis {$\{\Phi_0, \mathcal{P}^{(0)}, {\cal S}\}$.}

The transformation laws are
\begin{equation}\label{Tr-Low-new}
       \delta_\xi \Omega=   \tilde{T} \, , \qquad
      \delta_\xi \mathcal{P}^{(0)}= -\tilde{W} \, ,  \qquad
      \delta_\xi {\cal S} = \tilde{Y} \partial_\phi {\cal S}+2 {\cal S} \partial_\phi \tilde{Y} \, .
\end{equation}
Therefore, in this case, Poisson brackets \eqref{3d,d'}, \eqref{3d,e'} and \eqref{3d,f'} can be replaced by
\begin{equation}\begin{split}
\{{\cal S}(v,\phi), \Omega(v,\phi')\}&=\{{\cal S}(v,\phi), \mathcal{P}^{(0)}(v,\phi')\}=0\\
\{{\cal S}(v,\phi), {\cal S}(v,\phi')\}&= \left({\cal S}(v,\phi')\,\partial _\phi-{\cal S}(v,\phi)\,\partial_{\phi'}\right) \delta (\phi-\phi')\,.
%      [\bar\Upsilon_{m}(v), \bar\Upsilon_{n}(v') ] &= (m-n)\, \bar\Upsilon_{m+n}(v)\,\delta (v-v') \,\\
      \end{split}
\end{equation}
{In this basis, the charge algebra is Heisenberg $\oplus$ Diff($S^1)$, or Heisenberg $\oplus$ Witt. Due to its very simple form and the fundamental role this algebra plays in the analysis of change of basis, we call it fundamental NBS algebra.  In terms of Fourier modes and after ``quantization'' $i\,\{\,, \, \} \to [\, , \,]$, it takes the form
\begin{subequations}
    \begin{align}
        & [\Omega_m(v), \Omega_n(v) ] = [\mathcal{P}^{(0)}_{m}(v), \mathcal{P}^{(0)}_{n}(v) ] =0 \, , \label{s=0,b}\\
        & [\Omega_{m}(v), \mathcal{P}^{(0)}_{n}(v) ] = \frac{i}{8 G}\,\delta_{m+n,0}\,  , \label{s=-1,c}\\
        & [{\cal S}_{m}(v), {\cal S}_{n}(v) ] =(m-n)\, {\cal S}_{m+n}(v)\,  , \label{s=-1,d}\\
        & [{\cal S}_{m}(v), \Omega_{n}(v) ]=  [{\cal S}_{m}(v), \mathcal{P}^{(0)}_{n}(v) ]= 0 \, .\label{s=-1,f}
    \end{align}
\end{subequations}

The charges are explicitly integrable in the new basis and their algebra is clearly a Lie-algebra.

\subsubsection{Further integrable Lie-algebras} \label{sec:4.2.2}

The fundamental basis provides a suitable ground for exploring other changes of basis. Noting the direct sum structure of this algebra, we can readily observe that one may obtain the two parameters $\gamma,s$-family of algebras, where  the algebra becomes a direct sum of \eqref{2d-changed-basis-2} and Witt algebra.

While one can try to  exhaustively explore  the changes of basis from the fundamental algebra which keep the algebra {an} integrable Lie-algebra, we find it more useful and illuminating to discuss two interesting examples. {We will only comment on the issue of exhaustiveness at the end of the section}. 

To facilitate the analysis we first introduce two current algebras
\begin{equation}\label{def-j+-}
    J^\pm := \frac{1}{16\pi G} \left( \Omega \mp 2 G k \, \partial_\phi \mathcal{P}^{(0)} \right),
\end{equation}
with some constant  $k$.  The relation among chemical potentials is
\begin{equation}
    \tilde{W}=  \epsilon^+ + \epsilon^-  \, , \qquad \tilde{T}= 2 G k\, \partial_\phi \left( \epsilon^+ - \epsilon^- \right) \, .
\end{equation}
 Therefore, in this basis the charge variation is
\begin{equation}
\delta Q_\xi =  \int_0^{2\pi} \d \phi \left( \epsilon^+ \delta J^+ + \epsilon^- \delta J^- {+\tilde Y \delta \mathcal{S}} \right) \, .
\end{equation}
 The transformation laws \eqref{Tr-Low-new} now yields
\begin{equation}
    \delta_\xi J^\pm = \pm \frac{k}{4 \pi} \, \partial_\phi \epsilon^\pm \,, \end{equation}
and hence the charge algebra is given by
\begin{equation}\label{Currents+Witt}
\begin{split}
\{ J^\pm (v,\phi), J^\pm (v,\phi') \}&= \pm \frac{k}{4\pi} \, \partial_{\phi} {\delta} (\phi-\phi'), \qquad \{ J^\pm (v,\phi), J^\mp (v,\phi') \}=0.
\end{split}
\end{equation}

This current algebra is very similar to the one obtained as near horizon symmetries of BTZ black holes or $3d$ flat space cosmologies \cite{Afshar:2016kjj, Afshar:2016wfy}. There are, however, two crucial differences: 
\begin{itemize}
    \item [1] Our charges, besides $\phi$ dependence, are also $v$ dependent. The algebras there may be viewed at {a} constant $v$ slice of ours. 
    \item[2] Besides $J^\pm$ our NBS algebra has the ${\cal S}$ part, generating a Witt algebra, which is  absent in those  analyses. 
\end{itemize}

The first example is the {Vir} $\oplus$ {Vir} $\oplus$ Witt algebra. Given the $J^\pm, {\cal S}$ charges one can define a new basis as
\begin{equation}\label{L+-}
    L^\pm (v,\phi):=  \frac{2\pi}{k} \left[ J^\pm( v,\pm \phi)\right]^2 +{\beta_\pm} \partial_\phi J^\pm(v,\pm \phi)
\end{equation}
for which we have
\begin{equation}
    \epsilon^\pm(v,\phi) = \left(\frac{4\pi}{k}   J^\pm(v,\phi) \mp {\beta_\pm} \partial_\phi \right) \chi^\pm( v,\pm \phi) \, ,
    \end{equation}
where $\chi^\pm$ and $\tilde{Y}$ are chemical potentials in the new basis. The charge variation is
\begin{equation}
   \delta Q_{\xi}= \int_0^{2\pi} \d \phi \, (\chi^{+} \delta L^+ +\chi^{-} \delta L^- {+\tilde Y \delta \mathcal{S}})\, .
\end{equation}
The  variations of new generators $L^\pm$ are hence
\begin{equation}
    \delta_\xi L^\pm = \chi^\pm \partial_\phi L^\pm +2 L^\pm \partial_\phi \chi^\pm - \frac{k}{4\pi} {\beta_\pm^2}\partial_\phi^3 \chi^\pm \, .
\end{equation}
The complete charge algebra in this basis is
\begin{subequations}
    \begin{align}
    \{L^\pm (v,\phi), L^\pm (v,\phi')\}&= \left(L^\pm (v,\phi')\,\partial _\phi-L^\pm (v,\phi)\,\partial_{\phi'} + \frac{k}{4\pi} {\beta_\pm^2} \partial_{\phi'}^3 \right) {\delta} (\phi-\phi')\, , \label{LpmLpm}\\
    \{L^\pm (v,\phi), L^\mp (v,\phi')\}&=0 ,\qquad \{L^\pm (v,\phi), {\cal S} (v,\phi')\}= 0 \label{Lpm-S}\\
        \{ {\cal S}(v,\phi),  {\cal S}(v,\phi')\}&= \left( {\cal S}(v,\phi')\,\partial _\phi- {\cal S}(v,\phi)\,\partial_{\phi'}\right) {\delta} (\phi-\phi')\, .\label{SS-3d}
    \end{align}
\end{subequations}
This is a direct sum of three algebras and the central charge of the two Virasoro algebras $c_\pm=6 k \beta_\pm^2$ are arbitrary undetermined numbers.

The second example is the {BMS$_3\ \oplus$ Witt algebra.} Consider the two currents in the off-diagonal basis,
\begin{equation}\label{def-JK}
    J:= J^+ + J^-,\qquad K:= J^+- J^-
\end{equation}
and define 
\begin{equation}\label{L&M}
    L= \frac{{2}\pi}{k} J K + {\beta}\partial_\phi K+ {\alpha}\partial_\phi J\, , \hspace{0.7 cm} M= \frac{\pi}{k } J^2 +{\beta}\partial_\phi J \, .
   \end{equation}
The charge is then 
\begin{equation}
   \delta Q_{\xi}= \int_0^{2\pi} \d \phi \, (\epsilon_{\text{\tiny{L}}} \delta L +\epsilon_{\text{\tiny{M}}} \delta M {+\tilde Y \delta \mathcal{S}})\, , 
\end{equation}
with 
\begin{equation}
\epsilon^\pm = \frac{{2}\pi}{k} \left[ (K \pm J ) \epsilon_{\text{\tiny{L}}} +  \epsilon_{\text{\tiny{M}}} J\right] - \beta \partial_\phi (\epsilon_{\text{\tiny{M}}} \pm \epsilon_{\text{\tiny{L}}})-\alpha \partial_{\phi}\epsilon_{\text{\tiny{L}}}\, .
\end{equation}
The fields transform as
\begin{equation}
    \begin{split}
        \delta_\xi L=&\partial_\phi L \,  \epsilon_{\text{\tiny{L}}} + 2 L \partial_\phi \epsilon_{\text{\tiny{L}}} +\partial_\phi M \epsilon_{\text{\tiny{M}}} + 2 M \partial_\phi \epsilon_{\text{\tiny{M}}} - \frac{k}{2\pi} {\beta^2} \partial_\phi^3 \epsilon_{\text{\tiny{M}}} - \frac{k}{\pi} \alpha \beta \partial_\phi^3 \epsilon_{\text{\tiny{L}}}\, \\
         \delta_\xi M=& \partial_\phi M \epsilon_{\text{\tiny{L}}} + 2 M \partial_\phi \epsilon_{\text{\tiny{L}}} - \frac{k}{2\pi} {\beta^2} \partial_\phi^3 \epsilon_{\text{\tiny{L}}} \, , 
      \end{split}
\end{equation}
and the algebra is 
\begin{equation}
    \begin{split}
    \{M (v,\phi), M (v,\phi')\}=&0 \, , \\
         \{L (v,\phi), L (v,\phi')\}=& \left(L (v,\phi')\,\partial _\phi-L (v,\phi)\,\partial_{\phi'}+ \frac{k}{\pi} {\alpha\beta} \partial_{\phi'}^3  \right) {\delta} (\phi-\phi')\, , \\
         \{L (v,\phi), M (v,\phi')\}=& \left(M (v,\phi')\,\partial _\phi-M (v,\phi)\,\partial_{\phi'} + \frac{k}{2\pi} {\beta^2} \partial_{\phi'}^3 \right) {\delta} (\phi-\phi')\, , \\
    \end{split}
\end{equation}
which is a centrally extended BMS$_3$ algebra with central charges $c_{\text{\tiny{LM}}}=12k \beta^2, c_{\text{\tiny{LL}}}=24k \alpha\beta$ \cite{Barnich:2006av, Barnich:2011ct}. It is notable that the asymptotic symmetry algebra of $3d$ flat space over null infinity only realizes the $c_{\text{\tiny{LM}}}$ central charge \cite{Barnich:2006av, Barnich:2011ct}. itt part generated by ${\cal S}$.

These two examples are actually the only types of integrable Lie algebras that can be obtained from the Heisenberg sector of the fundamental basis without mixing the Heisenberg and Witt part.\footnote{A possibility of mixing the two sectors is given in \ref{sec:integrabe-3d-basis} where the algebra is a semidirect sum of Heisenberg and the Witt algebra generated by  $\Upsilon (v,\phi)$.} Let us start with the two currents $J^\pm$ or $J, K$. If the charge redefinitions involve powers of currents higher than two, the algebra would not close, we get something like $W_\infty$ algebras. Moreover, one can show that having more than one derivative of currents will not close either. Therefore, the expressions like \eqref{L+-} or \eqref{L&M} which are of twisted-Sugawara type\footnote{We note that within the class of twisted-Sugawara constructions, of course, one can obtain a bigger family of algebras than Vir $\oplus$ Vir or BMS$_3$: One can get all the family of centrally extended $W(a,b)$ algebras and in general, all algebraic deformations of centrally extended BMS$_3$, see  \cite{Parsa:2018kys} for more discussions.} \cite{Afshar:2016wfy} are the only options. On the other hand the Witt algebra of the fundamental basis cannot be deformed since it is stable/rigid \cite{Parsa:2018kys}.
%%%%%%%%%%%%%%%%%%%%%%%%%%%%%%%%%%%%%%%%%%%%%%%%%%%%%%%%%%%%%5
\section{Discussion and Outlook}\label{sec:conclusion}
%%%%%%%%%%%%%%%%%%%%%%%%%%%%%%%%%%%%%%%%%%%%%%%%%%%%%%%%%%%%%%'

In this work, we studied ``asymptotic symmetries'' near a null boundary in two and three-dimensional gravity theories. This work  is motivated by questions regarding black holes and is a continuation of similar analysis in four-dimensional vacuum Einstein theory \cite{Adami:2020amw}. We analysed the problem in full generality and in particular, we did not impose any specific boundary conditions. We were hence able to realize the maximal boundary degrees of freedom (b.d.o.f.), i.e. two charges which are arbitrary functions of light-cone time variable $v$ for $2d$ case and three charges with arbitrary dependence on $v,\phi$ coordinates in $3d$ case. Not imposing any boundary condition means that the theory which governs the b.d.o.f. is not specified and their $v$ dependence remains arbitrary. This is how our work differs from all previous works in the literature, except \cite{Adami:2020amw}. 

{In the literature the boundary conditions are typically specified by (1) falloff behaviour of the fields (here metric components in a specific coordinate system) and (2) choosing the field dependence of symmetry generators. As our analysis shows these two are conceptually independent and should be discussed separately. This second part is often not duly emphasised in the literature. To achieve the ``maximal b.d.o.f.'' analysis here we were forced not to impose restrictions on the falloff condition around $r=0$ null surface, we only assumed smoothness. We analysed the role of field dependence through the change of basis discussed in section \ref{sec:changebasis}. As our analysis explicitly demonstrates, the field dependence does not necessarily influence number of b.d.o.f.}

As our other important result,  we established that in $2d$ and $3d$ cases there exists a basis in which the charges are integrable. In other words, non-integrability of charges in $2d$ and $3d$ cases may be removed by working in a particular state/field-dependent bases. We discussed that the integrable basis is not unique.  We wrote the charges in different bases and computed their algebra. In $2d$ case we exhausted all such integrable Lie-algebra bases and in $3d$ case presented arguments what  these exhaustive bases are expected to be. It would, of course, be interesting to provide a rigorous proof of the latter.

Moreover, expanding upon the discussions and examples in \cite{Grumiller:2019fmp, Grumiller:2019ygj}, we  elaborated on the general state/field-dependent  change of basis which moves us in the family of integrable charges and also changes their algebra. We gave examples of how such change of basis can lead to non-zero central charges in  the charge algebra. That is, {the} existence of a central charge {in our maximal b.d.o.f analysis} can be an artifact of the basis used to present the charges. 

In the course of this paper, in the introduction and in section \ref{sec:changebasis} we have posed some questions but discussed and analysed a few of them. Here we would like to  expand further on those and discuss future projects and new directions. 

%\paragraph{Boundary conditions, variational principle, and the boundary/edge theory.} In our analysis, we only fixed a null surface to be sitting at $r=0$ and considered the most general expansion of metric around this surface which is allowed by field equations. We did not impose any boundary condition on the metric near the null surface. This is in contrast to the usual surface charge analysis where physically motivated boundary conditions are imposed through  {the} addition of boundary terms to the action and requiring the variational principle. As a result, our surface charges have arbitrary $v$ dependence. 

{\paragraph{Change of basis and field dependence of symmetry generators.}  In our examples we explored two classes of change of basis: Those which render a non-integrable charge integrable (\emph{cf}. discussions in sections \ref{sec:2dgrav} and \ref{sec:3dgrav}) and those which move us within the class of integrable charges, \emph{cf.} discussions in section \ref{sec:changebasis}. In both of these classes the change of bases are generically field-dependent.}

{We mentioned above that choice of the field dependence of symmetry generators is an essential part of fixing the boundary conditions and in both of these classes change of basis  should be viewed as choosing different boundary conditions. As the thermodynamical terminology we have adopted indicates, change of basis is conceptually analogous to change of ensemble in the thermodynamical systems. Each ensemble is specified by which quantities are held fixed while the associated conjugate thermodynamical charges are allowed to vary.}

{In a different viewpoint, our solution space may be viewed as a phase space where the charges are labelling the points in the solution phase space and the symmetry generators move us in this phase space. Explicitly, given the set of charges associated with the symmetry generators $\xi_i$, $Q_{\xi_i}$, $\delta_{\xi_i} Q_{\xi_j}=Q_{[\xi_i,\xi_j]}+ \text{central terms}$, where $Q_{[\xi_i,\xi_j]}$ labels a different point in the phase space.
A change of basis is then like a general coordinate transformation on this phase space (labelled by the charges). This change of basis in general specifies how we move on the phase space, it changes  the charge algebra.\footnote{{As a clarifying example, let us consider phase space of a given mechanical system in Hamiltonian formulation. Sympletomorphisms (``canonical transformations'') are a subset of possible change of basis which keep some given structures, the canonical Poisson brackets, intact. One may consider more general coordinate transformations on this phase space which also changes the basic Poisson brackets and takes us to a non-Barboux basis.}}}

{We also remark that in the both classes of our change of basis discussed above, in this paper we only restricted our analysis to change of basis within the maximal b.d.o.f. setting. It is, however, possible to impose further restrictions on these maximal charges ($d$ functions on $d-1$ surface, in the $d$ dimensional gravity setting) e.g. restrict them to $n$ ($n<d$) number of charges on $d-p$ ($p>1$) surfaces, together with our change of basis. We did not explore such possibilities in this work, but we expect to be able to impose all possible boundary conditions (in the sense discussed in the second paragraph of this section) starting from our maximal b.d.o.f setting.}

\paragraph{Boundary conditions and the variational principle.} In our analysis, we only fixed a null surface to be sitting at $r=0$ and considered the most general expansion of metric around this surface which is allowed by field equations. We did not impose any boundary condition on the metric near the null surface. This is in contrast to the usual surface charge analysis where physically motivated boundary conditions are imposed. 

Imposing variational principle, which may require addition of appropriate boundary terms, has a direct implication on the boundary conditions. In particular, it defines the boundary dynamics of the b.d.o.f, residing on the boundary.  This latter would restrict $v$ (time) dependence of our b.d.o.f. This procedure and imposing specific boundary conditions was carried out in $2d$ JT gravity \cite{Grumiller:2015vaa} and the AdS$_3$  \cite{Grumiller:2016pqb} and $3d$ flat \cite{Grumiller:2017sjh} cases.  This hence relates {to} our ``most general'' surface charges and those in these papers. 

%\tcb{Moreover,  a well defined variational principle relates the choice of the chemical potentials and  the choice of boundary conditions, and hence the choice of basis; it determines  which symmetry generators are kept fixed, see for e.g \cite{Grumiller:2019fmp}.} 
As our analysis and discussions implicitly imply,  the choice of boundary conditions and the charge basis are closely related to each other  and the information about these two is encoded in the chemical potentials. In analogy with usual thermodynamics, the choice of a boundary condition and a basis is like {the} choice of the ensemble which manifests itself through the chemical potentials which are variables conjugate to the (conserved) charges appearing in the first law. 

Determining the appropriate boundary theory at a null surface may give a better handle on {the} formulation of the membrane paradigm for black holes \cite{Thorne:1986iy, Parikh:1998mg}. It would be interesting to study the membrane paradigm by the addition of the surface degrees of freedom and how they may help with {the} identification of black hole microstates and resolution of the information puzzle. See \cite{Grumiller:2018scv} for preliminary discussions on this idea.

\paragraph{Modified bracket vs. integrable basis.} Non-integrable surface charges can generally appear in the covariant phase space analysis and in the literature so far two methods have been proposed to deal with them. (1) The Wald-Zoupas method \cite{Wald:1999wa} which prescribes adding a boundary term to the action to absorb the non-integrable (flux) part of the charge variation through the boundary. In this method a ``reference point'' in the phase space, for which we expect having a vanishing flux, is needed and may not be applied to generic cases, in particular to NBS analysis, see \cite{Adami:2020amw} for more discussions. (2) The Barnich-Troessaert modified bracket method \cite{Barnich:2011mi}, prescribing to separate the charge variation into an integrable part and a flux. The  ambiguity in this separation is then fixed by introducing a modified bracket such that algebra of the integrable part of the charges matches with the algebra of the symmetry generators, possibly up to a central extension; see appendix \ref{BT-MB-Appendix} for analysis of the modified bracket method for the $2d$ and $3d$ cases. 
%The Wald-Zoupas and Barnich-Troessaert modified bracket methods in cases when both apply lead to the same result. However, therein general may lead to different algebras. \cnote{Actually it is not clear. when they both of them exist they seem to agree. Do you have a counterexample?} 
%The former is more physical and gives a meaning to the flux while the latter is more mathematical and is based on algebraic cohomology intuitions. \cnote{the BT is also very physical - just the ambiguity is based on the bracket property while WZ heavily on diff geom. So maybe better to take it out }

In this work, we made the observation that non-integrability could be a result of {the} choice of basis for {the} charges and {the} corresponding chemical potentials. If an integrable basis exists,  we proved in section \ref{sec:changebasis} that such integrable bases are not unique. One may then ask if there always exists such integrable bases? Our analysis \cite{progress-2} shows that  the answer, in general, is no. In the particular case of $4d$ gravity one can show that, when there is a flux of propagating bulk d.o.f. (gravitons) through the boundary, the Bondi news, {it} will introduce ``genuine'' non-integrabilities which cannot be removed by a change of basis. This is of course expected: There is a close relation between genuine non-integrability and non-conservation of the charges, and {the} presence of non-trivial interactions between the boundary and bulk d.o.f  will render surface charges non-conversed and non-integrable. Making these statements more rigorous and robust and also connecting the ``change of basis method'' to Wald-Zoupas or Barnich-Troessaert modified bracket methods is postponed to our upcoming publication. 
%\cnote{Will be in one of two publications in the ref? }\snote{It can be :) but I do not want to commit it here.}

\paragraph{Higher-dimensional cases.} While we choose the lower dimensional gravity framework to establish our three main ideas, namely formulating the surface charge analysis in full generality near a null boundary, showing {the} existence of integrable basis and general change of basis among integrable charge Lie-algebras, our main goal is to apply this to $4d$ cases (black holes in real-world). This will enable us to extend the analysis in \cite{Adami:2020amw} to the most general case in which all the charges have arbitrary time dependence ($x^+$-dependence in the notation of that work) and in which the Bondi news through the null surface (horizon) is also allowed/turned on. This analysis will be presented elsewhere \cite{progress-1}. Here we discuss a partial result:

In the absence of Bondi news and ``genuine'' flux, the charges can be made integrable. In this case for generic dimension $d$, there is a ``fundamental NBS'' basis in which the algebra is Heisenberg $\oplus$ Diff$({\cal N}^{d-2}$), where ${\cal N}^{d-2}$ is the codimension two spacelike manifold  at constant $r,v$.  In the presence of genuine flux, one can get the same algebra among the integrable part of the charges, using the modified bracket method to separate the flux. 

Based on this algebra and for $d\geq 3$ one can construct BMS$_3\oplus {\cal D}$ where ${\cal D}$ is a subalgebra of Diff$({\cal N}^{d-2}$); see the analysis in section \ref{sec:4.2.2} for the $d=3$  case. This construction makes a closer connection to the near horizon BMS$_3$ algebras uncovered by Carlip \cite{Carlip:2017xne, Carlip:2019dbu}. 

It would be interesting to explore this line further and examine which boundary theories can be imposed on the system and how one can include more dynamical features like Bondi news, modeling matter falling into  or coming out of the black hole horizon {as} the Hawking radiation. 

%\cnote{I made a new paragraph }

%\snote{Hereafter to be edited.....}

%\paragraph{Symmetries on time-like or space-like surfaces and comparison to the null case.}

\section*{Acknowledgement}
We would like to especially thank Daniel Grumiller for his contribution in the earlier stages of this work and for many fruitful discussions in recent years  which led to this project and its development. We are grateful to Hamid Afshar and Hamid Safari  for their comments and discussions. VT would like to thank Mohammad Hassan Vahidinia for the useful discussions.
MMShJ  thanks the hospitality of ICTP HECAP  where a part of this research carried out. HY acknowledges Yau Mathematical Sciences Center for hospitality and support. HA acknowledges Saramadan grant No. ISEF/M/98204. MMShJ acknowledges the support by 
INSF grant No 950124 and Saramadan grant No. ISEF/M/98204.  The work of HY is supported in part by National Natural Science Foundation of China, Project 11675244. CZ was supported by the Austrian Science Fund (FWF), projects P 30822 and M 2665. 

\appendix

\section{Classification of Solutions to \emph{2d} Dilaton Gravity}\label{appen:2d-solutions}

The field equations for the action \eqref{action2d} are 
\bea 
&& \frac{\delta S}{\delta g_{\mu\nu}}=0\; \Rightarrow \; -\nabla_\mu\nabla_\nu\Phi+g_{\mu\nu}\nabla^2\Phi+\frac{1}{2}g_{\mu\nu}U=T_{\mu\nu}^{matter}\\
&& \frac{\delta S}{\delta \Phi}=0\; \Rightarrow \;  R-\frac{\d U}{\d\Phi}=0
\eea
In the absence of matter fields, one can solve equation of motion for a generic potential $U$.

The equation of motion for this case are
\be\label{EoM1}
R=\frac{\d U}{\d\Phi},\qquad 2\nabla_\mu\nabla_\nu\Phi+U g_{\mu\nu}=0,\qquad 
\ee 

A careful analysis reveals that there are two constant and linear dilaton cases:
\paragraph{Case I: Constant $\Phi$.} In this case the second equation in \eqref{EoM1} implies that this given constant $\Phi_0$ must be a zeros of the potential $U(\Phi_0)$. 
The Einstein equation then gives $R=\frac{dU}{d\Phi}|_{\Phi_0}=2K=Const.$ We get locally AdS$_2$, flat or dS$_2$ solutions corresponding to negative, zero or positive $K$. The most general solution can be obtained by a general coordinate transformation on the following metric
\be
\d s^2=\frac{2 \d x \d y}{\cosh^2 {\sqrt{\frac{K}{2}}}(x+y) }.
\ee 

Upon generic diffeomorphism a constant scalar remains a constant while the metric can become a constant curvature space in various coordinate systems.

\paragraph{Case II:  Linear $\Phi$.}
In $2d$ one can always bring any nonconstant scalar field $\Phi(x)$ to a linear function upon an appropriate coordinate transformation. So, we 
start with dilaton ansatz
\be\label{linear-dilaton}
 \Phi=l_\mu x^\mu
\ee
where $l_\mu=\partial_\mu\Phi$ can be a null, spacelike or timelike covector. The linear dilaton form \eqref{linear-dilaton} does not completely fix the coordinates, we can choose one of the coordinates $x$ such that $\Phi=x$ and the other coordinate $y$ may be chosen appropriately to make the metric ansatz simpler.

\begin{itemize}
    \item \textbf{$l^2\neq 0$ case.} We choose $y$ coordinate such that $g_{xy}=0,$ then equations of motion imply
\be\label{2d-sl-tl}
\d s^2= -\frac{1}{{\mathcal U}} \d x^2+  {\mathcal U} \d y^2,\qquad \Phi=x, \qquad \frac{\d{\mathcal U}}{\d\Phi}=U.
\ee 
\item \textbf{$l^2=0$ case.} We choose $y$ coordinate such that $g_{xx}=0$. The equations of motion, upon appropriate choice of the $y$ coordinate,  imply 
\be\label{2d-null}
\d s^2=2 \d x \d y +  {\mathcal U}\d y^2,\qquad \Phi=x, \qquad  \frac{\d{\mathcal U}}{\d\Phi}=U.
\ee
\end{itemize}
We note that in the linear dilaton basis, as \eqref{2d-sl-tl} and \eqref{2d-null} explicitly show, $\partial_y$ is a Killing vector. 

The most general solution is then obtained upon applying a generic diffeomorphism $(x, y)\to (\Phi, \Psi), \Phi=\Phi(x,y), \Psi=\Psi(x,y)$ on the above solutions. This general solution is hence specified by two arbitrary functions. The Killing vector for a generic solution is  $k=\epsilon^{\mu\nu}\partial_\mu\Phi\partial_\nu$. 
%%%%%%%%%%%%%%%%%%%%%%%%%%%%%%%%%%%%%%%%%%%%%%%%%%%

For our analysis an appropriate choice of coordinates is such that $r=0$ is a null surface:
\be\label{general-2d-metric}
\d s^2=2 \eta \d v \d r- F(v,r) \d v^2, \qquad F|_{r=0}=0.
\ee
For this choice equations of motion imply that 
\be
\partial_r^2\Phi=0\quad \Rightarrow \quad \Phi=\Phi_0(v) +\Phi_1(v) \; r
\ee
For $U(\phi)\neq 0$ case and when $\Phi, \Phi_0$ are not constant we get

\be\label{general-2d-solution}
F=-\eta\left(\frac{2 \Phi_1'}{\Phi_1}  r+\frac{ {\mathcal U}(\Phi_0+r\Phi_1)-{\mathcal U}(\Phi_0) }{\Phi_1^2} \eta \right),\qquad \Phi_1=-\frac{{\mathcal U}(\Phi_0)}{2\Phi_0'}   \eta,\qquad U=\frac{\d {\mathcal U}}{\d\Phi}.
\ee
In this case the solution is specified by two arbitrary functions $\eta (v), \Phi_0(v)$.

\paragraph{Example:} 
 A simple example of a theory with action \eqref{action2d} may be obtained by dimensional reduction from four dimensions. Consider the Einstein-Maxwell theory given by
\be
S=\frac{1}{16\pi G}\int \textrm{d} ^4x\sqrt{-h}\left(R_4-G {\cal F}^2\right) 
\ee 
with an ansatz for the magnetically charged solution as
\be\label{2d-RN-reduction-ansatz}
ds^2=\Phi^{-1/2} g_{\mu\nu}\d x^\mu \d x^\nu + G \Phi \d\Omega_2^2,\quad {\cal F}=p\sin\theta \d\theta \wedge \d\phi
\ee 
It is easy to check that equations of motion for $g_{\mu\nu}$ and $\Phi$ can be obtained from following two-dimensional action
\be\label{2d-MRN-action}
S=\frac{1}{4\pi}\int \textrm{d} ^2x \sqrt{-g}\left(\Phi R+\frac{2}{G \Phi^{1/2}}-\frac{2p^2}{G\Phi^{3/2}}\right)\,.
\ee

For theory in \eqref{2d-MRN-action} the solution which is obtained from reduction of 4d magnetically RN black hole solution over the $S^2$
takes the form
\be
g_{\mu\nu}\d x^\mu \d x^\nu=G{\rho}\left(-f \d t^2+\frac{\d\rho^2}{f}\right) ,\quad \Phi={\rho^2}
\ee 
where 
\be
f=\frac{(\rho-\rho_+)(\rho-\rho_-)}{\rho^2},\qquad \rho_\pm=M\pm\sqrt{M^2 -p^2}
\ee 
To rewrite metric into the form \eqref{general-2d-solution} we take
\be
\rho =\sqrt{2r+\rho_+^2},\qquad dv={G}(\d t+\frac{\d\rho}{f})
\ee 
we get
\be
g_{\mu\nu}\d x^\mu \d x^\nu=2\d v\d r-F \d v^2 ,\quad \Phi=\rho_+^2+{2r}
\ee 
where
\be F=\frac{\left(\sqrt{2r+\rho_+^2}-\rho_+\right)\left(\sqrt{2 r+\rho_+^2}-\rho_-\right)}{G\sqrt{2  r+\rho_+^2}}\ee
Expanding $F$ for small $r$ we get
\be F=\frac{\rho_+-\rho_-}{G\rho_+^2} r +{\mathcal O}(r^2)
\ee 
Finally, one may build a general two function family of solutions upon the coordinate transformations 
\be\label{2d-diffeos-r,v}
r\to r\;\frac{\eta}{V'},\qquad v\to V,
\ee
where $\eta, V$ are two arbitrary functions of $v$ and $V'=dV/dv$.

\section{Classification of AdS$_3$ Solutions}\label{appen:3d-solutions}

As a starter let us consider with a BTZ black hole solution \cite{Banados:1992wn} in the BTZ coordinates
\be
\d s^2=-g \d t^2 + \frac{ \d \rho^2}{g}+\rho^2\left(\d \psi-N \d t\right)^2,\quad g=\frac{(\rho^2-r_+^2)(\rho^2-r_-^2)}{\ell^2 \rho^2},\quad N=\frac{r_+r_-}{\ell \rho^2}
\ee
with $\psi\equiv \psi+2\pi$. The surface gravity $\kappa$ and horizon angular velocity $\Omega$ for the above solutions are
\be
\kappa= \frac{r_+^2 -r_-^2}{\ell^2 r_+},\qquad \Omega_{\text{\tiny H}}= \frac{r_-}{ \ell r_+}\,.
\ee
Then defining 
\be
\rho= r+r_+,\quad \d t=\d v-\frac{\d r}{g},\quad \d \psi=\d \phi + \Omega_{\text{\tiny H}} \d v -\frac{N \d r}{g} 
\ee
bring the BTZ metric into the Gaussian Null Coordinate (GNC), 
\be
\d s^2 = - 2 \, \kappa \, r \left( 1+ \frac{r}{2\, r_+}\right) \d v^2+ 2 \d v \d r + \frac{4 \, r \, r_-}{\ell} \left( 1+ \frac{r}{2\, r_+}\right) \d v \d \phi + (r_{+} + r)^2 \d \phi^2
\ee 
where $\partial_v$ is a Killing vector generating the horizon which is now sitting at $r=0$. One can check that for this metric
$g^{rr}|_{r=0}=0$. 

The coordinate transformations\footnote{An interesting method which leads to a specific three function family of solutions is to promote horizon radii $r_\pm$ to functions. That is,
\be\label{r+-}
r\to \rho=r\eta(v,\varphi),\qquad r_\pm\to r_\pm(v,\varphi),
\ee
where $r_\pm(v,\varphi+2\pi)=r_\pm(v,\varphi)$. The family in \eqref{r+-} yields a solution only if $r_\pm$ have a specific $v,\varphi$ dependence: 
\be
r_\pm(v,\varphi)= {\cal J}_+(x^+)\pm {\cal J}_-(x^-),
\ee
where $x^\pm= v/l \pm \varphi$. For $L_\pm(x^\pm)={\cal J}^2_\pm(x^\pm)$, this is the family of solutions discussed  in \cite{Troessaert:2013fma}. }
\be\label{transformation1}
r\to \frac{\eta}{\ell V_{,v}} \, r ,\quad v\to \ell V,\quad \phi\to\Phi
\ee
with 
\be 
\eta(v,\varphi+2\pi)=\eta(v,\varphi),\quad V(v,\varphi+2\pi)=V(v,\varphi), \quad \Phi(v,\varphi+2\pi)=\Phi(v,\varphi)+2\pi,
\ee
are the most general ones which keep $r=0$ a null surface. Under coordinate transformations \eqref{transformation1} we obtain a three-function family of solutions
\bea\label{metric1}
&&
\d s^2=2\eta \d v \d r -\left(F_0+F_1 \frac{r}{\ell} +F_2\frac{r^2}{\ell^2}\right)\d v^2 +2\ell\left(f_0+f_1 \frac{r}{\ell} +f_2\frac{r^2}{\ell^2}\right)\d v\d\phi\nonumber \\ &&\qquad +\frac{2\eta\partial_\phi V}{\partial_v V}\d r\d\phi +\ell^2\left(h_0+h_1 \frac{r}{\ell}+h_2\frac{r^2}{\ell^2}\right)\d\phi^2
\eea
where
\begin{subequations}
\begin{align}
& h_0=\left(\frac{r_+\partial_\phi \Phi}{\ell}\right)^2 \\
&h_1=\frac{2\partial_\phi\eta\partial_\phi V }{\ell\partial_v V }-\frac{2r_+ \eta (\partial_\phi V )^2}{\ell^2\partial_v V }+\frac{2\eta (r_+ \partial_\phi \Phi +r_-\partial_\phi V )^2}{\ell^2 {\, r_+ \,}\partial_v V }-\frac{2\eta\partial_\phi V {\partial_{v} \partial_\phi V} }{\ell (\partial_v V )^2} \\ 
& h_2=\left[\frac{\eta \left(r_+\partial_\phi  \Phi +r_- \partial_\phi V \right) }{ {\ell r_+} \partial_v V }\right]^2-\left(\frac{\eta\partial_\phi  V }{\ell\partial_v V }\right)^2\\
&F_0=-\left(r_+\partial_v \Phi\right)^2\\ 
&F_1=2r_+\eta\partial_v V -2\ell \partial_v\eta+\frac{2\ell\eta\partial_v^2 V }{\partial_v V }-\frac{2\eta (r_+ \partial_v \Phi +r_-\partial_vV )^2}{r_+\partial_v V }
\\ & F_2=\eta^2\left[1- \left(\frac{{r_-}\partial_vV+ {r_+}\partial_v \Phi }{{r_+}\partial_v V } \right)^2 \right]
\\ & f_0=\frac{r_+^2 \partial_v \Phi \partial_\phi \Phi}{\ell} 
\\ &f_1= \partial_\phi\eta -\frac{\eta\partial_{v} \partial_\phi V }{\partial_v V }-\frac{2r_+\eta {\partial_\phi V}}{\ell}+\frac{\partial_\phi V \partial_v\eta}{\partial_v V }-\frac{\eta\partial_\phi V \partial_v^2 V }{(\partial_v V )^2}\nonumber \\ &\qquad +\frac{2\eta (r_+\partial_v \Phi +r_-\partial_v V)(r_+\partial_\phi \Phi +r_-\partial_\phi V) }{\ell {r_+} \partial_v V }
\\ & f_2= \frac{\eta^2 }{\ell(\partial_v V )^2} \left(\frac{\left(r_+\partial_\phi\Phi +r_-\partial_\phi V \right)\left(r_+\partial_v\Phi +r_- \partial_v V \right)}{r_+^2}- \partial_\phi V \partial_v V\right)
\end{align}
\end{subequations}

For solution \eqref{metric1}, the horizon Killing vector is 
\be
\xi=\frac{{\partial_\phi\Phi}}{{\ell\left(\partial_\phi\Phi\partial_v V-\partial_v\Phi\partial_\phi V\right)}}\left(\partial_v +r \left(\frac{\partial_v^2V}{\partial_v V}-\frac{\partial_v\Phi\partial_{v\phi}^2V}{\partial_vV\partial\phi\Phi}-\frac{\partial_v\eta}{\eta}\right) \partial_r-\frac{\partial_v\Phi}{\partial_\phi\Phi}\partial_\phi\right)
\ee 
It is easy to check that 
\be|\xi|^2=0\qquad \Rightarrow \qquad r=0, r=-\frac{2 r_+\ell \partial_v V}{\eta} \ee
which shows $r=0$ is the horizon, as it is expected from the coordinate transformation \eqref{r+-}.

{One can also solve the Einstein equations directly to get the complete solution. In order to do that, first we use three diffeomorphisms to set $g_rr=g_{r\phi}=0$, $g^{rr}|_{r=0}=0$, and $\partial_r\eta=0$. As it has been discussed in \cite{Adami:2020amw}, there is no surface charge associate to a diffeomorphism which brings back r-dependence to $\eta$.} 

Starting from ansatz \eqref{3d-NH-metric}, the $rr$ component we can find $r$ dependent of $h$
\be
{\mathcal E}_{rr}=0\quad\Rightarrow \quad h=\left(\Omega +\frac{h_1}{2\Omega}\; r\right)^2
\ee
where $\Omega$ and $h_1$ are functions of $v$ and $\phi$. Using this, $r\phi$ component of the Einstein equations can be solve and it determines $r$ dependent of $f$ as follows
\be
{\mathcal E}_{r\phi}=0\quad\Rightarrow \quad f=f_0 +f_1 r + \frac{(f_1-\partial_\phi\eta)h_1}{4\Omega^2}\; r^2
\ee
where $f_0$ and $f_1$ are functions of $v$ and $\phi$. Then $\phi\phi$ component of the Einstein can be solved and it fixes $r$ dependent of $F$, 
\be
{\mathcal E}_{\phi\phi}=0\quad\Rightarrow \quad F=F_0 +F_1 r + F_2 r^2
\ee
where
\bea
&& F_1=\frac{4 \eta^2 \Omega^2}{\ell^2 h_1} +\frac{2\eta\partial_\phi f_1-f_1^2 -2\eta\partial_\phi^2\eta-2\eta\partial_v h_1+\partial_\phi\eta^2}{h_1} \nonumber \\ && \hspace{8mm} - \frac{2\eta(f_1\partial_\phi\Omega-\partial_\phi \eta\partial_\phi\Omega-h_1\partial_v\Omega)}{\Omega h_1}\\
&& F_2=\frac{ \eta^2}{\ell^2} -\frac{(f_1-\partial_\phi\eta)^2}{4\Omega^2}
\eea
where $\Lambda=-\frac{1}{\ell^2}$. One can check that the with above solution, $vr$ component of the Einstein equation is satisfied itself. Therefore we left with two more equations ${\mathcal E}_{vv}={\mathcal E}_{v\phi}=0$ to solve for six functions $\eta, \Omega, h_1, f_0, f_1$ and $F_0$ of $v$ and $\phi$.

To solve the rest of the equations we first note that under following coordinate transformation
\be\label{cte}  \phi\to\alpha(v,\phi) ,\qquad r\to \frac{r}{\eta} +\beta(v,\phi),\ee and appropriate choice for $\alpha$ and $\beta$ we can set

 $F_0, f_0\to 0$ and $\eta\to 1$.  We use this fact to set  $F_0=f_0=0$ and $\eta=1$ first.   Equation of motion for the rest of metric components are simplified significantly now and we can solve them easily.  After solving all equations and finding the solution, we retrieve functions $F_0, f_0$ and $\eta$ using a coordinate transformation like \eqref{cte}. Up to two functions of $\phi$, appear as integration constants and can be absorbed by proper scaling later, solving equations of motion gives 
\bea
 &&h_1=\frac{\ell^4-4\partial_\phi\Omega+(3\partial_\phi\Omega^2-2\Omega\partial_\phi^2\Omega) \ell^2+\Omega^4}{\ell^2\Omega \partial_v\Omega} \\
 && f_1=\frac{2\partial_\phi\Omega}{\Omega}-\frac{2\partial_{v\phi}^2\Omega   }{\partial_v\Omega}-\frac{2\ell}{\Omega}
 \eea
 Then after a coordinate transformation of form  \eqref{cte}, we retrieve $f_0, F_0$ and $\eta$. The final solution depends on four functions $\alpha, \beta, \eta$ and $\Omega$. Furthermore imposing $g^{rr}|_{r=0}$ condition, eliminates one free function from the solution and gives a family of solutions with three independent functions. 

We expect a similar analysis can be carried out for $3d$ flat and dS$_3$ cases.

\section{Modified Bracket and Integrable Parts of Charges}\label{BT-MB-Appendix}

Given a non-integrable charge variation, one can always decompose it as
\begin{equation}\label{BT-separation}
    \slashed\delta Q_\xi=\delta Q^I_\xi + \mathcal{F}_{\xi}
\end{equation}
where $Q^I_\xi$ is the integrable part of the charge and $\mathcal{F}_{\xi}$ is the non-integrable part of the charge, the flux. There is an ambiguity in the above separation which may be fixed
by the Barnich-Troessaert modified bracket  \cite{Barnich:2011ct, Barnich:2011mi}, defined as
\begin{equation}\label{MB-definition}
    \begin{split}
        \left\{ Q_{\xi_1}^{\text{I}} , Q_{\xi_2}^{\text{I}}\right\}^{*}:=  Q_{[\xi_{1},\xi_2]}^{\text{I}} \equiv \delta_{\xi_2} Q_{\xi_1}^{\text{I}} + \mathcal{F}_{\xi_{2}}[\delta_{\xi_1}\psi].  
         \end{split}
\end{equation}
where $\psi$ are the fields parameterising our solution space. 
One may solve \eqref{MB-definition} to fix  $\delta Q^I$ and hence the flux ${\cal F}$. Below, we do so for the $2d$ and $3d$ examples. See \cite{Adami:2020amw} for similar analysis for the $4d$ case. 

\paragraph{Modified bracket and the $2d$ example.} We start with a decomposition of \eqref{Charge-variation-01} as in \eqref{BT-separation}, 
\begin{subequations}\label{charge-flux-2d-flat}
\begin{align}
   Q_\xi^{\text{I}}= & \frac{1}{16 \pi G} \left[ W \, \Phi_0 -  T \left(\Gamma\, \Phi_0  +  2\,\Phi_0^{\prime}  \right)  \right] \label{Integrable-part} \\ 
    \mathcal{F}_\xi=& \frac{1}{16 \pi G} \left( \Phi_0 \delta \Gamma  + \frac{\Phi_0' \, \delta  \eta}{\eta}  \right) T \, ,
\end{align}
\end{subequations}
This decomposition is of course not unique and  there is an ambiguity,
\begin{equation}
    \tilde{Q}^{\text{I}}_{\xi}= Q_{\xi}^{\text{I}}+ \mathcal{A}_\xi [\psi]\, , \qquad \tilde{\mathcal{F}}_\xi = \mathcal{F}_\xi -\delta \mathcal{A}_\xi [\psi]\, ,  
\end{equation}
where $\psi\in\{\Phi_0, \eta\}$. 
The above ambiguity will be fixed by \eqref{MB-definition} up to a  shift in the central extension term
\begin{equation}\label{CET-01}
    K_{\xi_1, \xi_2}= \delta_{\xi_2} \mathcal{A}_{\xi_1} - \delta_{\xi_1} \mathcal{A}_{\xi_2} - \mathcal{A}_{[\xi_1,\xi_2]}.
\end{equation}
To see this let us assume,
\begin{equation}
    \mathcal{A}_\xi [\psi]= \mathcal{P}[\psi] \, T + \mathcal{R}[\psi] \, W \, ,
\end{equation}
where $\mathcal{P}$ and $\mathcal{R}$ are two functionals of $\psi$. The highest derivative of $T$ allowed to appear in $K_{\xi_1, \xi_2}$ is three. Then, the highest derivatives of $\eta$ and $\Phi_0$ allowed to appear in $\mathcal{P}$ and $\mathcal{R}$ are respectively two and three. Demanding  the algebra of surface charges to be isomorphic to the algebra of NBS vector fields \eqref{2d-algebra}, up to a central extension term, one finds that $\mathcal{R}=0$ and $\mathcal{P}$ must satisfy following conditions
\begin{subequations}\label{CET-02}
    \begin{align}
        & \frac{\partial \mathcal{P}}{\partial \eta}=\frac{\partial \mathcal{P}}{\partial \eta'}=\frac{\partial \mathcal{P}}{\partial \eta''}=0 \, ,\\
        & \Phi_0'\frac{\partial \mathcal{P}}{\partial \Phi_0''} + 3 \Phi_0''\frac{\partial \mathcal{P}}{\partial \Phi_0'''}=0 \, , \\
        & \Phi_0'\frac{\partial \mathcal{P}}{\partial \Phi_0'}+ 2 \Phi_0''\frac{\partial \mathcal{P}}{\partial \Phi_0''} + 3 \Phi_0'''\frac{\partial \mathcal{P}}{\partial \Phi_0'''}=\mathcal{P} \, .
    \end{align}
\end{subequations}
The central extension term \eqref{CET-01} subject to \eqref{CET-02}, can be rewritten as
\begin{equation}\label{CET-03}
    K_{\xi_1, \xi_2}= \Phi_0'\frac{\partial \mathcal{P}}{\partial \Phi_0'''}\left( T_1 T_2''' - T_2 T_1''' \right)
\end{equation}
which ought to be a constant for BMS$_3$ algebra \eqref{BMS3-2d-KVA}. This fact, together with \eqref{CET-02}, leads to $\mathcal{P}= \lambda(\Phi_0) \, \Phi_0'$ and hence $K_{\xi_1, \xi_2}=0$. That is, our charge and flux decomposition has fixed the ${\cal A}_\xi$ ambiguity up to ${\cal A}_\xi=\lambda(\Phi_0) \Phi_0' T.$ This, however, does not change the charge algebra which is by construction BMS$_3$ \eqref{BMS3-2d-KVA}. Explicitly, the Laurent expansion modes of the integrable part of the charges are
\be\label{Int-charge-2d-flat}
{\cal W}_n:=\frac{1}{2\pi i}\oint \frac{v^{n+1}}{16 \pi G} \Phi_0  \, \qquad {\cal T}_n:=\frac{1}{2\pi i}\oint \frac{v^{n+1}}{16 \pi G} \left( 2\,\Phi_0^{\prime} +\Phi_0 \,\Gamma \right)
\ee
which satisfy the BMS$_3$ algebra
\be\label{charge-algera-2d-flat}
\begin{split}
[{\cal T}_n, {\cal T}_m] & = (n-m){\cal T}_{n+m}\\
[{\cal W}_n, {\cal W}_m] & = 0\\ 
[{\cal T}_n, {\cal W}_m] & = (n-m){\cal W}_{n+m}
\end{split}\ee
We also remark that using the definition of charges and flux above one may obtain the ``generalized charge conservation equation'' (GCCE) \cite{Adami:2020amw}
\be\label{GCCE-2d-flat}
\partial_v Q_{\xi}^{I} =-\mathcal{F}_{\partial_{v}}(\delta_{\xi}\psi).
\ee

\paragraph{Modified bracket and the $3d$ example.}
One can use the same procedure to split the $3d$ charge variation \eqref{charge-variation''}  into its integrable and non-inetgrable parts. Since the analysis is quite similar to the $2d$ case we skip the computations and only present the final result for the decomposition satisfying the modified bracket condition \eqref{MB-definition}:
\begin{subequations}\label{charge-flux-decomposition-3d}
\begin{align}
   Q_\xi^{\text{I}} = & \frac{1}{16 \pi G} \int_{0}^{2\pi} d \phi \left\{ W \Omega + Y \Upsilon + T \left[ -2 \chi - \Omega\, \Gamma + \frac{f_0 \, \Upsilon}{\Omega^2} + \partial_\phi \left( \frac{f_0}{\Omega}\right)\right]\right\} \label{3d-Integrable-part} \\ 
    \mathcal{F}_\xi (\delta g)= & \frac{1}{16 \pi G} \int_{0}^{2\pi} d \phi \, T\left\{  \Omega \delta \Gamma - \Upsilon \delta \left(\frac{ f_0}{\Omega^2}\right) - \partial_\phi \left[ \Omega \delta \left(\frac{ f_0}{\Omega^2}\right)\right] + \frac{ \chi\delta \eta}{\eta}\right\} \, ,\label{3d-flux-part} 
\end{align}
\end{subequations}
where $\delta g$ stands for a generic variation of the metric. As a check of computations, one may then verify that $Q_\xi^I$ satisfy the algebra \eqref{3d-NBS-KV-algebra} without any central charge. Moreover, one can observe the GCCE, $\partial_v Q_{\xi}^{I} =-\mathcal{F}_{\partial_{v}}(\delta_{\xi}{g}),$ also holds in the $3d$ case.

\bibliographystyle{fullsort.bst}

\providecommand{\href}[2]{#2}\begingroup\raggedright\endgroup

%\bibliography{reference}

\begin{thebibliography}{10}

\bibitem{Grumiller:2020vvv}
D.~Grumiller, M.~Sheikh-Jabbari, and C.~Zwikel, ``{Horizons 2020},''
  \href{http://www.arXiv.org/abs/2005.06936}{{\tt 2005.06936}}.

\bibitem{Sheikh-Jabbari:2016lzm}
M.~M. Sheikh-Jabbari, ``{Residual diffeomorphisms and symplectic soft hairs:
  The need to refine strict statement of equivalence principle},'' {\em Int. J.
  Mod. Phys.} {\bf D25} (2016), no.~12, 1644019,
\href{http://www.arXiv.org/abs/1603.07862}{{\tt 1603.07862}}.
%%CITATION = ARXIV:1603.07862;%%.

\bibitem{Bondi:1962}
H.~Bondi, M.~van~der Burg, and A.~Metzner, ``Gravitational waves in general
  relativity {VII.} {W}aves from axi-symmetric isolated systems,'' {\em Proc.
  Roy. Soc. London} {\bf A269} (1962) 21--51.

\bibitem{Sachs:1962}
R.~Sachs, ``Asymptotic symmetries in gravitational theory,'' {\em Phys. Rev.}
  {\bf 128} (1962) 2851--2864.

\bibitem{strominger:2017zoo}
A.~Strominger, ``{Lectures on the Infrared Structure of Gravity and Gauge
  Theory},'' \href{http://www.arXiv.org/abs/1703.05448}{{\tt 1703.05448}}.

\bibitem{Lee:1990nz}
J.~Lee and R.~M. Wald, ``{Local symmetries and constraints},'' {\em J. Math.
  Phys.} {\bf 31} (1990)
725--743.
%%CITATION = JMAPA,31,725;%%.

\bibitem{Iyer:1994ys}
V.~Iyer and R.~M. Wald, ``Some properties of {N}{\"o}ther charge and a proposal
  for dynamical black hole entropy,'' {\em Phys. Rev.} {\bf D50} (1994)
  846--864,
\href{http://arXiv.org/abs/gr-qc/9403028}{{\tt gr-qc/9403028}}.
%%CITATION = GR-QC 9403028;%%.

\bibitem{Compere:2018aar}
G.~Compère and A.~Fiorucci, ``{Advanced Lectures on General Relativity},''
  {\em Lect. Notes Phys.} {\bf 952} (2019) 150,
\href{http://www.arXiv.org/abs/1801.07064}{{\tt 1801.07064}}.
%%CITATION = ARXIV:1801.07064;%%.

\bibitem{Oblak:2016eij}
B.~Oblak, {\em {BMS Particles in Three Dimensions}}.
\newblock PhD thesis, Brussels U., 2016.
\newblock \href{http://www.arXiv.org/abs/1610.08526}{{\tt 1610.08526}}.

\bibitem{Wald:1999wa}
R.~M. Wald and A.~Zoupas, ``{A General definition of 'conserved quantities' in
  general relativity and other theories of gravity},'' {\em Phys.Rev.} {\bf
  D61} (2000) 084027,
\href{http://www.arXiv.org/abs/gr-qc/9911095}{{\tt gr-qc/9911095}}.
%%CITATION = GR-QC/9911095;%%.

\bibitem{Barnich:2011mi}
G.~Barnich and C.~Troessaert, ``{BMS charge algebra},'' {\em JHEP} {\bf 1112}
  (2011) 105,
\href{http://www.arXiv.org/abs/1106.0213}{{\tt 1106.0213}}.
%%CITATION = ARXIV:1106.0213;%%.

\bibitem{Barnich:2010eb}
G.~Barnich and C.~Troessaert, ``{Aspects of the BMS/CFT correspondence},'' {\em
  JHEP} {\bf 1005} (2010) 062,
\href{http://www.arXiv.org/abs/1001.1541}{{\tt 1001.1541}}.
%%CITATION = ARXIV:1001.1541;%%.

\bibitem{Brown:1986nw}
J.~D. Brown and M.~Henneaux, ``{Central Charges in the Canonical Realization of
  Asymptotic Symmetries: An Example from Three-Dimensional Gravity},'' {\em
  Commun. Math. Phys.} {\bf 104} (1986)
207--226.
%%CITATION = CMPHA,104,207;%%.

\bibitem{Thorne:1986iy}
K.~S. Thorne, R.~Price, and D.~Macdonald, {\em Black Holes: The Membrane
  Paradigm}.
\newblock Yale University Press,
1986.
\newblock
%%CITATION = ISBN-9780300037708 ETC.;%%.

\bibitem{Parikh:1998mg}
M.~K. Parikh, ``{Membrane horizons: The Black hole's new clothes},'' other
  thesis, Princeton University, 10, 1998.

\bibitem{Hawking:2016msc}
S.~W. Hawking, M.~J. Perry, and A.~Strominger, ``{Soft Hair on Black Holes},''
  {\em Phys. Rev. Lett.} {\bf 116} (2016), no.~23, 231301,
\href{http://www.arXiv.org/abs/1601.00921}{{\tt 1601.00921}}.
%%CITATION = ARXIV:1601.00921;%%.

\bibitem{Donnay:2015abr}
L.~Donnay, G.~Giribet, H.~A. Gonz{\'a}lez, and M.~Pino, ``{Supertranslations
  and Superrotations at the Black Hole Horizon},'' {\em Phys. Rev. Lett.} {\bf
  116} (2016), no.~9, 091101,
\href{http://www.arXiv.org/abs/1511.08687}{{\tt 1511.08687}}.
%%CITATION = ARXIV:1511.08687;%%.

\bibitem{Donnay:2016ejv}
L.~Donnay, G.~Giribet, H.~A. Gonz{\'a}lez, and M.~Pino, ``{Extended Symmetries
  at the Black Hole Horizon},'' {\em JHEP} {\bf 09} (2016) 100,
\href{http://www.arXiv.org/abs/1607.05703}{{\tt 1607.05703}}.
%%CITATION = ARXIV:1607.05703;%%.

\bibitem{Afshar:2016kjj}
H.~Afshar, D.~Grumiller, W.~Merbis, A.~Perez, D.~Tempo, and R.~Troncoso,
  ``{Soft hairy horizons in three spacetime dimensions},'' {\em Phys. Rev.}
  {\bf D95} (2017), no.~10, 106005,
\href{http://www.arXiv.org/abs/1611.09783}{{\tt 1611.09783}}.
%%CITATION = ARXIV:1611.09783;%%.

\bibitem{Afshar:2016wfy}
H.~Afshar, S.~Detournay, D.~Grumiller, W.~Merbis, A.~Perez, D.~Tempo, and
  R.~Troncoso, ``{Soft Heisenberg hair on black holes in three dimensions},''
  {\em Phys. Rev.} {\bf D93} (2016), no.~10, 101503,
\href{http://www.arXiv.org/abs/1603.04824}{{\tt 1603.04824}}.
%%CITATION = ARXIV:1603.04824;%%.

\bibitem{Afshar:2016uax}
H.~Afshar, D.~Grumiller, and M.~M. Sheikh-Jabbari, ``{Near horizon soft hair as
  microstates of three dimensional black holes},'' {\em Phys. Rev.} {\bf D96}
  (2017), no.~8, 084032,
\href{http://www.arXiv.org/abs/1607.00009}{{\tt 1607.00009}}.
%%CITATION = ARXIV:1607.00009;%%.

\bibitem{Mao:2016pwq}
P.~Mao, X.~Wu, and H.~Zhang, ``{Soft hairs on isolated horizon implanted by
  electromagnetic fields},''
\href{http://www.arXiv.org/abs/1606.03226}{{\tt 1606.03226}}.
%%CITATION = ARXIV:1606.03226;%%.

\bibitem{Grumiller:2016kcp}
D.~Grumiller, A.~Perez, S.~Prohazka, D.~Tempo, and R.~Troncoso, ``{Higher Spin
  Black Holes with Soft Hair},'' {\em JHEP} {\bf 10} (2016) 119,
\href{http://www.arXiv.org/abs/1607.05360}{{\tt 1607.05360}}.
%%CITATION = ARXIV:1607.05360;%%.

\bibitem{Grumiller:2018scv}
D.~Grumiller and M.~M. Sheikh-Jabbari, ``{Membrane Paradigm from Near Horizon
  Soft Hair},'' {\em Int. J. Mod. Phys.} {\bf D27} (2018) 1847006,
\href{http://www.arXiv.org/abs/1805.11099}{{\tt 1805.11099}}.
%%CITATION = ARXIV:1805.11099;%%.

\bibitem{Ammon:2017vwt}
M.~Ammon, D.~Grumiller, S.~Prohazka, M.~Riegler, and R.~Wutte, ``{Higher-Spin
  Flat Space Cosmologies with Soft Hair},'' {\em JHEP} {\bf 05} (2017) 031,
\href{http://www.arXiv.org/abs/1703.02594}{{\tt 1703.02594}}.
%%CITATION = ARXIV:1703.02594;%%.

\bibitem{Chandrasekaran:2018aop}
V.~Chandrasekaran, {\'E}.~{\'E}. Flanagan, and K.~Prabhu, ``{Symmetries and
  charges of general relativity at null boundaries},'' {\em JHEP} {\bf 11}
  (2018) 125,
\href{http://www.arXiv.org/abs/1807.11499}{{\tt 1807.11499}}.
%%CITATION = ARXIV:1807.11499;%%.

\bibitem{Chandrasekaran:2019ewn}
V.~Chandrasekaran and K.~Prabhu, ``{Symmetries, charges and conservation laws
  at causal diamonds in general relativity},''
\href{http://www.arXiv.org/abs/1908.00017}{{\tt 1908.00017}}.
%%CITATION = ARXIV:1908.00017;%%.

\bibitem{Grumiller:2019fmp}
D.~Grumiller, A.~Pérez, M.~Sheikh-Jabbari, R.~Troncoso, and C.~Zwikel,
  ``{Spacetime structure near generic horizons and soft hair},'' {\em Phys.
  Rev. Lett.} {\bf 124} (2020), no.~4, 041601,
  \href{http://www.arXiv.org/abs/1908.09833}{{\tt 1908.09833}}.

\bibitem{Adami:2020amw}
H.~Adami, D.~Grumiller, S.~Sadeghian, M.~Sheikh-Jabbari, and C.~Zwikel,
  ``{T-Witts from the horizon},'' {\em JHEP} {\bf 04} (2020) 128,
  \href{http://www.arXiv.org/abs/2002.08346}{{\tt 2002.08346}}.

\bibitem{Jackiw:1984}
R.~Jackiw, ``{Liouville field theory: A two-dimensional model for gravity?},''
  in {\em Quantum Theory Of Gravity}, S.~Christensen, ed., pp.~403--420.
\newblock Adam Hilger, Bristol, 1984.

\bibitem{Teitelboim:1984}
C.~Teitelboim, ``{The {H}amiltonian structure of two-dimensional space-time and
  its relation with the conformal anomaly},'' in {\em Quantum Theory Of
  Gravity}, S.~Christensen, ed., pp.~327--344.
\newblock Adam Hilger, Bristol, 1984.

\bibitem{Callan:1992rs}
C.~G. Callan, Jr., S.~B. Giddings, J.~A. Harvey, and A.~Strominger,
  ``Evanescent black holes,'' {\em Phys. Rev.} {\bf D45} (1992) 1005--1009,
\href{http://www.arXiv.org/abs/hep-th/9111056}{{\tt hep-th/9111056}}.
%%CITATION = HEP-TH 9111056;%%.

\bibitem{Brown:1988am}
J.~Brown, {\em {LOWER DIMENSIONAL GRAVITY}}.
\newblock World Scientific, 1988.

\bibitem{Grumiller:2002nm}
D.~Grumiller, W.~Kummer, and D.~Vassilevich, ``{Dilaton gravity in
  two-dimensions},'' {\em Phys. Rept.} {\bf 369} (2002) 327--430,
  \href{http://www.arXiv.org/abs/hep-th/0204253}{{\tt hep-th/0204253}}.

\bibitem{Grumiller:2015vaa}
D.~Grumiller, J.~Salzer, and D.~Vassilevich, ``{AdS$_{2}$ holography is
  (non-)trivial for (non-)constant dilaton},'' {\em JHEP} {\bf 12} (2015) 015,
\href{http://www.arXiv.org/abs/1509.08486}{{\tt 1509.08486}}.
%%CITATION = ARXIV:1509.08486;%%.

\bibitem{Grumiller:2017qao}
D.~Grumiller, R.~McNees, J.~Salzer, C.~Valcárcel, and D.~Vassilevich,
  ``{Menagerie of AdS$_{2}$ boundary conditions},'' {\em JHEP} {\bf 10} (2017)
  203, \href{http://www.arXiv.org/abs/1708.08471}{{\tt 1708.08471}}.

\bibitem{Gegenberg:1994pv}
J.~Gegenberg, G.~Kunstatter, and D.~Louis-Martinez, ``Observables for
  two-dimensional black holes,'' {\em Phys. Rev.} {\bf D51} (1995) 1781--1786,
\href{http://www.arXiv.org/abs/gr-qc/9408015}{{\tt gr-qc/9408015}}.
%%CITATION = GR-QC 9408015;%%.

\bibitem{Barnich:2006av}
G.~Barnich and G.~Compere, ``{Classical central extension for asymptotic
  symmetries at null infinity in three spacetime dimensions},'' {\em
  Class.Quant.Grav.} {\bf 24} (2007) F15--F23,
\href{http://www.arXiv.org/abs/gr-qc/0610130}{{\tt gr-qc/0610130}}.
%%CITATION = GR-QC/0610130;%%.

\bibitem{Barnich:2011ct}
G.~Barnich and C.~Troessaert, ``{Supertranslations call for superrotations},''
  {\em PoS} {\bf CNCFG2010} (2010) 010,
  \href{http://www.arXiv.org/abs/1102.4632}{{\tt 1102.4632}}.
[Ann. U. Craiova Phys.21,S11(2011)].
%%CITATION = ARXIV:1102.4632;%%.

\bibitem{Compere:2015knw}
G.~Comp{\`e}re, P.-J. Mao, A.~Seraj, and M.~M. Sheikh-Jabbari, ``{Symplectic
  and Killing symmetries of AdS$_{3}$ gravity: holographic vs boundary
  gravitons},'' {\em JHEP} {\bf 01} (2016) 080,
\href{http://www.arXiv.org/abs/1511.06079}{{\tt 1511.06079}}.
%%CITATION = ARXIV:1511.06079;%%.

\bibitem{Banados:1992gq}
M.~Ba\~nados, M.~Henneaux, C.~Teitelboim, and J.~Zanelli, ``Geometry of the
  (2+1) black hole,'' {\em Phys. Rev.} {\bf D48} (1993) 1506--1525,
\href{http://www.arXiv.org/abs/gr-qc/9302012}{{\tt gr-qc/9302012}}.
%%CITATION = GR-QC 9302012;%%.

\bibitem{Banados:1992wn}
M.~Ba\~nados, C.~Teitelboim, and J.~Zanelli, ``The black hole in
  three-dimensional space-time,'' {\em Phys. Rev. Lett.} {\bf 69} (1992)
  1849--1851,
\href{http://www.arXiv.org/abs/hep-th/9204099}{{\tt hep-th/9204099}}.
%%CITATION = HEP-TH 9204099;%%.

\bibitem{Cornalba:2002fi}
L.~Cornalba and M.~S. Costa, ``{A New cosmological scenario in string
  theory},'' {\em Phys.Rev.} {\bf D66} (2002) 066001,
\href{http://www.arXiv.org/abs/hep-th/0203031}{{\tt hep-th/0203031}}.
%%CITATION = HEP-TH/0203031;%%.

\bibitem{Cornalba:2003kd}
L.~Cornalba and M.~S. Costa, ``{Time dependent orbifolds and string
  cosmology},'' {\em Fortsch.Phys.} {\bf 52} (2004) 145--199,
\href{http://www.arXiv.org/abs/hep-th/0310099}{{\tt hep-th/0310099}}.
%%CITATION = HEP-TH/0310099;%%.

\bibitem{Bagchi:2012xr}
A.~Bagchi, S.~Detournay, R.~Fareghbal, and J.~Simon, ``{Holography of 3d Flat
  Cosmological Horizons},'' {\em Phys. Rev. Lett.} {\bf 110} (2013) 141302,
\href{http://www.arXiv.org/abs/1208.4372}{{\tt 1208.4372}}.
%%CITATION = ARXIV:1208.4372;%%.

\bibitem{Spradlin:2001pw}
M.~Spradlin, A.~Strominger, and A.~Volovich, ``{Les Houches lectures on de
  Sitter space},'' in {\em {Les Houches Summer School: Session 76: Euro Summer
  School on Unity of Fundamental Physics: Gravity, Gauge Theory and Strings}},
  pp.~423--453.
\newblock 10, 2001.
\newblock \href{http://www.arXiv.org/abs/hep-th/0110007}{{\tt hep-th/0110007}}.

\bibitem{Parsa:2018kys}
A.~Farahmand~Parsa, H.~R. Safari, and M.~M. Sheikh-Jabbari, ``{On Rigidity of
  3d Asymptotic Symmetry Algebras},'' {\em JHEP} {\bf 03} (2019) 143,
\href{http://www.arXiv.org/abs/1809.08209}{{\tt 1809.08209}}.
%%CITATION = ARXIV:1809.08209;%%.

\bibitem{gao2011low}
S.~Gao, C.~Jiang, and Y.~Pei, ``Low-dimensional cohomology groups of the lie
  algebras w (a, b),'' {\em Communications in Algebra{\textregistered}} {\bf
  39} (2011), no.~2, 397--423.

\bibitem{progress-2}
H.~Adami, D.~Grumiller, M.~Sheikh-Jabbari, V.~Taghiloo, H.~Yavartanoo, and
  C.~Zwikel, ``On integrability of surface charges at null boundaries,'' {\em
  Work in preparation} (2020).

\bibitem{Safari:2019zmc}
H.~R. Safari and M.~M. Sheikh-Jabbari, ``{BMS$_{4}$ algebra, its stability and
  deformations},'' {\em JHEP} {\bf 04} (2019) 068,
\href{http://www.arXiv.org/abs/1902.03260}{{\tt 1902.03260}}.
%%CITATION = ARXIV:1902.03260;%%.

\bibitem{Grumiller:2019ygj}
D.~Grumiller, M.~Sheikh-Jabbari, C.~Troessaert, and R.~Wutte, ``{Interpolating
  Between Asymptotic and Near Horizon Symmetries},'' {\em JHEP} {\bf 03} (2020)
  035, \href{http://www.arXiv.org/abs/1911.04503}{{\tt 1911.04503}}.

\bibitem{Grumiller:2016pqb}
D.~Grumiller and M.~Riegler, ``{Most general AdS$_{3}$ boundary conditions},''
  {\em JHEP} {\bf 10} (2016) 023,
  \href{http://www.arXiv.org/abs/1608.01308}{{\tt 1608.01308}}.

\bibitem{Grumiller:2017sjh}
D.~Grumiller, W.~Merbis, and M.~Riegler, ``{Most general flat space boundary
  conditions in three-dimensional Einstein gravity},'' {\em Class. Quant.
  Grav.} {\bf 34} (2017), no.~18, 184001,
  \href{http://www.arXiv.org/abs/1704.07419}{{\tt 1704.07419}}.

\bibitem{progress-1}
H.~Adami, D.~Grumiller, M.~Sheikh-Jabbari, V.~Taghiloo, H.~Yavartanoo, and
  C.~Zwikel, ``News and twitts from the horizon,'' {\em Work in preparation}
  (2020).

\bibitem{Carlip:2017xne}
S.~Carlip, ``{Black Hole Entropy from Bondi-Metzner-Sachs Symmetry at the
  Horizon},'' {\em Phys. Rev. Lett.} {\bf 120} (2018), no.~10, 101301,
\href{http://www.arXiv.org/abs/1702.04439}{{\tt 1702.04439}}.
%%CITATION = ARXIV:1702.04439;%%.

\bibitem{Carlip:2019dbu}
S.~Carlip, ``{Near-Horizon BMS Symmetry, Dimensional Reduction, and Black Hole
  Entropy},''
\href{http://www.arXiv.org/abs/1910.01762}{{\tt 1910.01762}}.
%%CITATION = ARXIV:1910.01762;%%.

\bibitem{Troessaert:2013fma}
C.~Troessaert, ``{Enhanced asymptotic symmetry algebra of $AdS$$_{3}$},'' {\em
  JHEP} {\bf 08} (2013) 044,
\href{http://www.arXiv.org/abs/1303.3296}{{\tt 1303.3296}}.
%%CITATION = ARXIV:1303.3296;%%.

\end{thebibliography}
\end{document}